# Beyond variance: sensitivity-based dimensions in brain networks underlie individual differences in cognitive ability

**Sida Chen[1,2], Siqi Yang[1,2], Zhao Chang[1,2] , Taro Toyoizumi[3], Werner Sommer[4,7,8], Lianchun Yu[5,6*], Qian-Yuan Tang[1,3*], and Changsong Zhou[1,2,7*]**

[1] Department of Physics, Hong Kong Baptist University, Kowloon Tong, Hong Kong
[2] Centre for Nonlinear Studies, Hong Kong Baptist University, Kowloon Tong, Hong Kong
[3] Lab for Neural Computation and Adaptation, RIKEN Center for Brain Science, 2-1 Hirosawa, Wako, Saitama 351-0106, Japan
[4] Department of Psychology, Humboldt-Universitaet zu Berlin, Germany
[5]School of Physical Science and Technology, Lanzhou Center for Theoretical Physics, Lanzhou University, Lanzhou, Gansu 730000, China
[6] Key Laboratory of Theoretical Physics of Gansu Province, and Key Laboratory of Quantum Theory and Applications of MoE, Lanzhou University, Lanzhou, Gansu 730000, China
[7] Life Science Imaging Centre, Hong Kong Baptist University, Kowloon Tong, Hong Kong
[8] Faculty of Education, National University of Malaysia, Kuala Lumpur, Malaysia

## Abstract

A central goal of human neuroscience is to identify brain–behavior associations that are not only predictive, but also interpretable, reproducible and biologically grounded across scales. This goal remains difficult because behaviorally relevant individual differences are embedded within high-dimensional and coupled patterns of brain activity, connectivity and anatomy. Here we introduce a sensitivity-based network model that represents each individual as a deviation from a group-level brain network and ranks coordinated changes in regional excitability and interregional coupling by their consequences for whole-brain dynamics. In working-memory and language fMRI data from more than 1,000 participants, we found that cognitive individual differences were concentrated in a small subset of high-sensitivity eigennetworks, whereas much inter-individual variation lay in low-sensitivity dimensions. The resulting task-processing network was sparse, topologically coherent and stable across repeated resampling and reduced training samples, outperforming edge-wise connectome models in both prediction and feature reproducibility. Sensitivity-based sparsification showed that removing low-sensitivity variation preserved behavioral explanation, while disrupting high-sensitivity structure rapidly degraded it. The same network linked behavior to functional connectivity, structural connectivity, cortical microstructure and transcriptomic gradients. These results indicate that reproducible brain–behavior associations can be recovered by organizing individual differences according to their system-level consequences for brain dynamics and behavior, rather than according to the neural features that vary most across individuals.

## Significance / Cover Letter

Individual differences in cognitive abilities are a defining feature of human behavior, yet the largest inter-individual differences in brain networks need not be the ones that matter most for cognition. Here we introduce Stiff Network Modeling, a sensitivity-based framework that decomposes large-scale brain networks into eigennetworks organized by their behavioral impact rather than by inter-individual variance alone. Using fMRI data collected during working-memory and language tasks in a large participant cohort, we identified a topologically coherent brain network pattern whose coordinated variation best explains individual differences in cognitive task performance. Importantly, the same network-level representation provides a cross-scale link between functional dynamics, structural connectivity, cortical microstructure, and transcriptomic gradients. These findings suggest that individual cognitive abilities are organized along specific low-variance, high-impact network dimensions. Stiff Network Modeling provides a mechanistically interpretable framework for brain–behaviour association across scales.

## Introduction

Individual differences in cognitive abilities are a defining feature of human behaviour and are strongly linked to educational, occupational and health outcomes(Deary et al. 2010).. At the same time, large-scale brain organization is neither wholly shared nor wholly idiosyncratic: individualized network mapping and precision functional atlases show that canonical functional systems are preserved at the population level while displaying robust person-specific topography that relates to cognition(Bassett et al. 2017, Hermosillo et al. 2024, Keller et al. 2023).. A central challenge for human neuroscience is therefore not simply to detect brain–behaviour associations, but to identify neural systems that explain individual differences, remain stable across samples and can be related across scales. Recent brain-wide association studies have made this challenge explicit: reproducible effects are often small, sample-size requirements are high, and study design features such as scan duration and sampling scheme can strongly influence replicability and prediction accuracy(Kang et al. 2024, Marek et al. 2022, Ooi et al. 2025). Recent work further suggests that generalizable brain–behaviour associations may depend on identifying neural features that are stable at the individual level rather than simply prominent at the group level(Rodriguez et al. 2025, Tian et al. 2026). The problem is therefore not whether inter-individual neural variation exists, but which coordinated forms of brain variation are consequential for cognition.This challenge is sharpened by the organization of the brain itself. Contemporary network neuroscience has established that cognition is supported by distributed neural activity patterns and coordinated interactions across large-scale networks rather than isolated brain regions. Individual differences are therefore embedded in high-dimensional and strongly coupled patterns of activity, connectivity and anatomy. As a result, the neural features that vary most across individuals, distinguish individuals most strongly, or covary most strongly with behaviour in a given sample need not be the features that best explain how brain systems support cognition.(Luppi et al. 2026, Rodriguez et al. 2025). This distinction matters because brain–behaviour studies are increasingly expected to satisfy three demands at once: interpretability, reproducibility and biological grounding across scales.

Existing approaches each address part of this problem.Pattern-based methods such as multivoxel pattern analysis (MVPA) demonstrated that behaviorally relevant information is embedded in distributed activity patterns (Haxby et al. 2001, Norman et al. 2006; Fig. 1A), and connectome-based predictive modeling (CPM) extended this logic to distributed connectivity patterns (Shen et al. 2017; Fig. 1B). These approaches can yield intuitive maps of regions or edges, but when features are highly correlated, selected weights and feature sets may reflect sample-specific covariance structure rather than the neural components that are most consequential for behaviour(Haufe et al. 2014, Woo et al. 2017). More generally, latent-variable and generative approaches, including canonical correlation analysis (CCA), variational autoencoders (VAE) and latent-space statistical models, seek low-dimensional representations of brain–behavior relationships (Fig. 1C). Yet the dimensions they identify are often harder to relate back to concrete neural systems or interpretable model parameters. More generally, statistical association, individual discriminability and behavioural relevance are not interchangeable properties of brain variation(Mantwill et al. 2022, Rodriguez et al. 2025). What is therefore needed is not simply a more efficient low-dimensional representation, but one that can identify coordinated neural differences according to their consequences for brain dynamics and behaviour.

We propose that this problem can be addressed by shifting the focus from variation to consequence. If each individual is defined as a deviation from a shared group-level brain network, the key question becomes how coordinated deviations from that reference alter whole-brain activity in ways that matter for behaviour. This requires a representation that is both interpretable and system-level.Here we describe brain dynamics in terms of regional excitability and interregional coupling, and rank coordinated parameter directions by their influence on model-predicted brain activity using the Fisher information matrix. This logic is closely related to the stiff–sloppy organization of complex systems. In such systems, a small number of coordinated parameter combinations have strong effects on system output, whereas many others can vary substantially but with relatively little functional consequence (Gutenkunst et al. 2007, Machta et al. 2013, Transtrum et al.; Panas et al. 2015; Ponce-Alvarez et al. 2020, 2022). Related recent work has likewise shown that behaviourally relevant neural dynamics can be separated from other components of high-dimensional population activity, reinforcing the need to organize brain variation by consequence rather than by covariance alone(Sani et al. 2024).Guided by this idea, we introduce Stiff Network Modeling, which expresses each participant as a deviation from a group-level reference network, decomposes that parameter space into sensitivity-ranked eigennetworks and identifies behaviour-aligned combinations of those eigennetworks. (Fig. 1D). Applying this approach to large-scale fMRI data from working-memory and language tasks, we show that behaviorally relevant individual differences concentrate in a small subset of high-sensitivity eigennetworks, whereas much inter-individual variance is distributed across low-sensitivity dimensions. By combining the behaviorally relevant eigennetworks, we derive a task-processing network. The resulting task-processing network is sparse, topologically coherent and stable across repeated resampling and reduced training samples, and it links behaviour to functional connectivity, structural connectivity, cortical microstructure and transcriptomic gradients. Together, these results suggest that interpretable and reproducible brain–behaviour associations emerge when individual differences are organized by their consequences for brain dynamics and behaviour, rather than by the features that vary most across individuals.

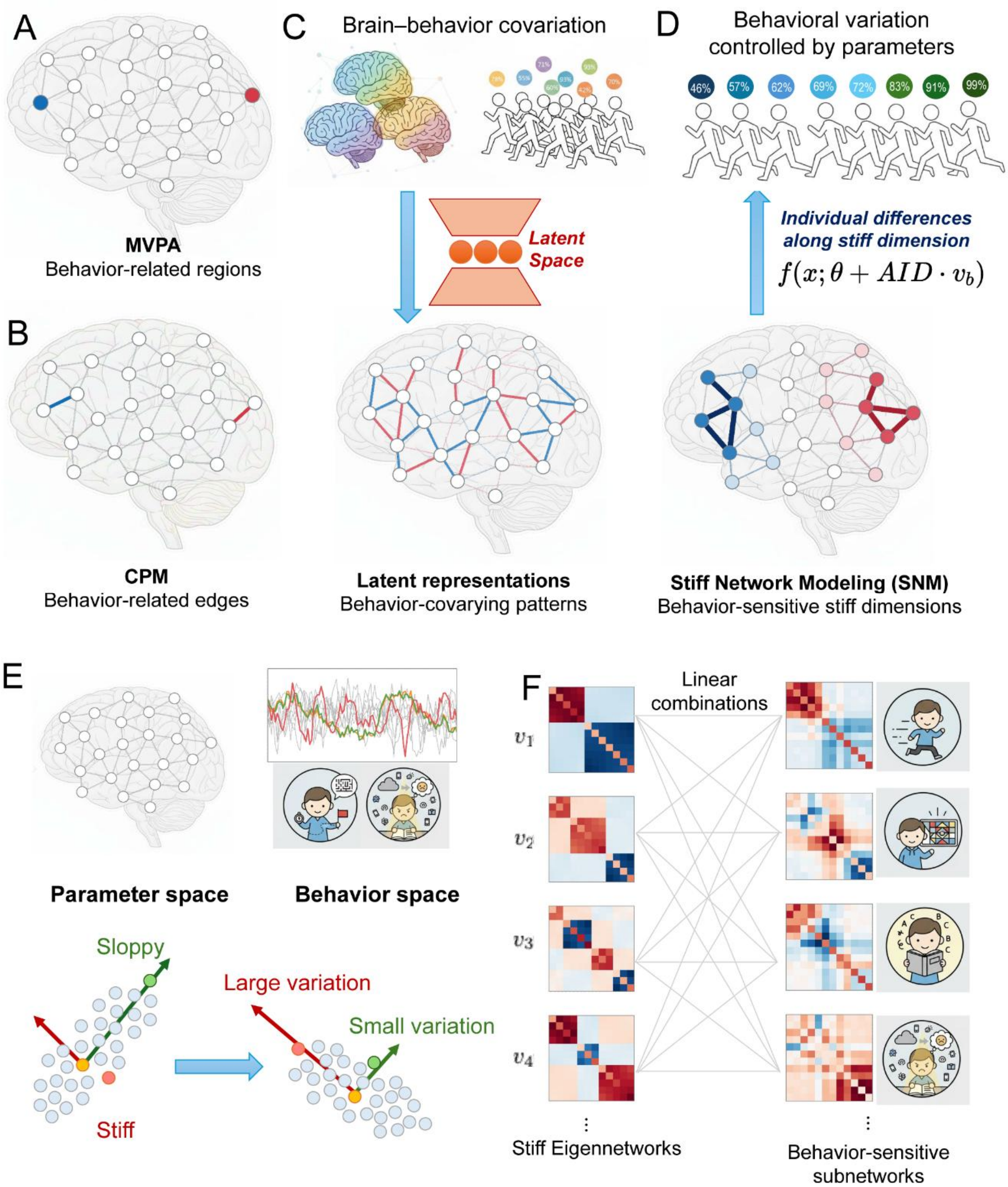

**Figure 1 | A schematic progression from covariation analysis to behavior-sensitive eigennetworks. (A)** Multivariate pattern analysis (MVPA) relates behavior to distributed brain activity by fitting contributions from individual regions and combining them at the pattern level. The interactions among regions are not explicitly taken into consideration by treating regional activity as independent indicators. **(B)** Connectome-based predictive modeling (CPM) extends this logic to connectivity by selecting and combining individual network edges whose average strengths predict behavioral performance. However, brain network links cannot be treated as independent indicators, as networks operate in coordinated modes. **(C)** Top-down latent and generative models represent brain–behavior relationships through low-dimensional axes that capture dominant patterns of covariation in brain and performance. These covariation modes reside in an abstract latent space, making their underlying network mechanisms difficult to interpret in terms of concrete system

parameters such as regional excitability or interregional coupling. **(D)** The Stiff Network Modeling (SNM), introduced here, adopts a sensitivity-based modeling approach by organizing network configurations along interpretable network parameter dimensions ranked by behavioral sensitivity when parameters change in coordinated manner along these dimensions. It isolates stiff directions in interpretable parameter space, corresponding to coordinated combinations of regional excitability and interregional coupling (red: positive sensitivity; blue: negative sensitivity), along which small perturbations (simultaneously increased coupling (red) and reduced coupling (blue)) induce large changes in task performance. We write the model output as $f(x;,\theta + \mathrm{AID} \cdot v_b)$, where x denotes the brain state, θ the group-level reference parameter vector, $v_b$ a behavior-related stiff direction (a coordinated combination of parameters), and AID a scalar coordinate that quantifies an individual's signed displacement along $v_b$ from θ. Importantly, these stiff directions need not coincide with directions of large inter-individual parameter variation or covariation. **(E)** Conceptual illustration of parameter space, where nodes represent regional excitability parameters and edges represent interregional coupling parameters that will be obtained by fitting a model to brain dynamical patterns. Parameter differences do not generally translate proportionally into behavioral differences: small deviations along stiff directions (orange vs. red) can produce large changes in brain activity (illustrated by time series) and cognitive performance, whereas large deviations along sloppy directions (orange vs. green) have minimal behavioral impact. **(F)** SNM decomposes large-scale brain network into sensitivity-ranked eigennetworks $v_1, v_2, \ldots$. By optimization in identifying and selectively combining stiff eigennetworks, SNM constructs behavior-sensitive subnetworks for different tasks, targeting parameter directions (network combinations) most relevant to individual task performance rather than those exhibiting the largest statistical variability.

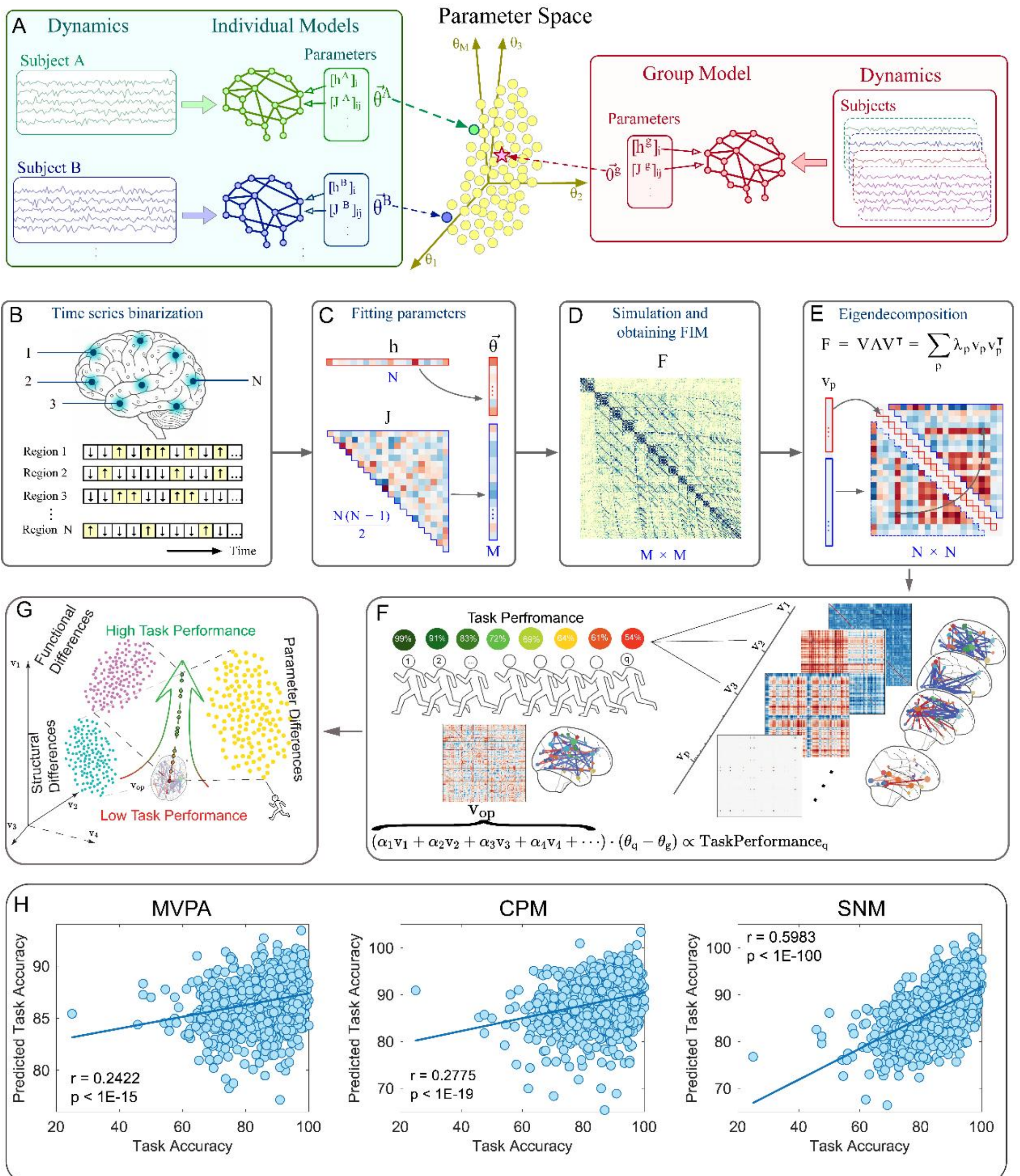

**Figure 2 | Steps of Stiff-Network Modeling and comparison with conventional predictive models.**

**(A-G)** Workflow of SNM. **(A)** Illustration of brain network parametrization by calibration of individual models (left) and group model (right) of brain activities. By fitting a pairwise maximum entropy model (PMEM) to binarized brain dynamics data (e.g., fMRI), variable brain dynamics states (more precisely, the probability distribution of state patterns) are transformed into static model parameters. The red star indicates the group parameters fitted to the group data. Each circle refers to the parameters fitted to a specific participant $q$ (e.g., A or B). Each parameter vector contains two kinds of parameters, $h$ and $J$, corresponding to regional excitability and interregional coupling, respectively. In this way, individual differences are analyzed in parameter space. **(B)** Normalizing and binarizing fMRI signals over time with a suitable threshold into up (+1) and down (−1) states (arrows). Black dots denote the centers of regions of interest (ROIs). **(C)** Fitting PMEM parameters. In parameter space $\vec{\theta} = (h, J)$, each model is represented by an $M$-dimensional parameter vector, containing the $N$-dimensional external field $h$ (related to regional excitability) and the *N(N−1)/2*-dimensional interregional coupling *J*. **(D)** Calculation of the Fisher information matrix (FIM). **(E)** Eigendecomposition of the FIM yields eigenvalues and eigenvectors (sensitivity-ranked directions). Each element of eigenvector $\overrightarrow{v_{\mathrm{op}}}$ corresponds one-to-one with model parameters $\vec{\theta} = (h, J)$, allowing elements to be reshaped into an $N \times N$ symmetric matrix, with $h$-related components on the diagonal and *J*-related components off-diagonal. These sensitivity-ranked eigenvectors thus correspond to interpretable eigennetworks along which coordinated parameter variation has impact on the global brain states quantified by the eigenvalues. **(F)** Eigennetwork decomposition and optimal combination. Right: Eigennetwork decomposition and visualization. Representative eigennetworks (e.g., top-ranked eigenvectors sorted by descending eigenvalue) derived from the group-level FIM on the Schaefer-100 parcellation are shown in matrix/network form. Blue and red links denote negative and positive sensitivities, respectively; node colors indicate the seven canonical Yeo functional networks (Yeo et al. 2011). Left: Optimal combination of eigennetworks to identify the task-processing network. Recruitment weights $\alpha_p$ are optimized across eigennetworks to define the optimal behavior-aligned direction $\overrightarrow{v_{\mathrm{op}}}$ corresponding to the task-processing network; projection of each individual's parameter-deviation vector from the group-level reference onto $\overrightarrow{v_{\mathrm{op}}}$ yields the *Behavior Aligned Individual Difference* (Behavior-AID), which best predicts cognitive scores. **(G)** Unified stiff-sloppy coordinate systems spanned by the sensitivity-ranked directions for individual differences in brain structure, function and cognitive performance, allowing the study of structure-function-behavior relationship in an integrative approach. Individuals (yellow points) form a cloud in eigennetwork coordinates in parameter space and are projected onto the behavior-relevant axis (dashed line). Points on the axis are ordered to best correlate with task performance (arrow color gradient from low to high). The schematic further links the ordered axis to individual variability in structural space (purple cloud; e.g., structural connectivity) and functional space (blue cloud; e.g., functional connectivity), illustrating that the weighted task-processing network encoded by $\overrightarrow{v_{\mathrm{op}}}$ extracts behavior-relevant neural components (combinations of regions and/or connections) in structural or functional properties. **(H)** Comparison of predictive modelling frameworks: MVPA (multi-variate pattern analysis, left) uses regional average BOLD signals as features, CPM (connectome-predictive modeling, middle) uses mean strength of task-related functional connections, and SNM (right) uses Behavior-AID derived from optimized task-processing networks. SNM provides both enhanced prediction accuracy and more interpretable network-level features (shown later). The blue solid lines show the least-squares fitting, with Pearson's correlation coefficient *r* and corresponding *p*-value indicated in each panel. The working memory task performance is used here.

## Results

### A sensitivity-based coordinate system reveals behaviorally relevant directions of brain variation

To examine why inter-individual neural variance does not map straightforwardly onto behavior, we first expressed individual brains in a common parameter space and ranked its coordinated directions by their impact on brain dynamics rather than by variance across individuals. We extracted task-fMRI BOLD time series (working memory and language) from 100 cortical regions defined by the Schaefer-100 atlas (Schaefer et al. 2018), aligned to the canonical seven functional networks (Yeo et al. 2011), and converted each regional signal into binary up/down states (Fig. 2A,B). We then fitted a pairwise maximum-entropy model (PMEM) (Watanabe et al. 2013) to the binarized multivariate activity patterns, transforming the empirical distribution of variable brain states into a static parameter vector ($\vec{\theta}$) (Fig. 2A,C, see Method). Each participant's model parameters $\overrightarrow{\theta_q}$ were represented by regional effective excitability terms $h$ and interregional coupling terms $J$. A group-level reference $\overrightarrow{\theta_g}$ was obtained by fitting PMEM to the data pooled from all participants, and individual brains were mapped onto the parameter space (Fig. 2A). We assessed fitting convergence at both group and individual PMEMs, and discard individuals without satisfactory fitting convergence (see Method). Across the retained participants, PMEM fitting trajectories showed stable convergence of model-simulated functional connectivity (FC) toward empirical FC (Supplementary Fig. S1A). From the group PMEM, we obtained the group FIM (Fig. 2D) and its corresponding eigennetworks from eigendecompostion (Fig. 2E, see Method). Repeated 3-fold analyses further showed high overlap among fold-specific leading eigenspaces and strong overlap between fold-specific and full-sample FIMs (Supplementary Fig. S1B), supporting the use of the full-sample group PMEM and FIM as a stable reference for sensitivity estimation. Eigennetworks were ordered by descending eigenvalues (rank 1 = stiffest), thereby defining a sensitivity hierarchy in parameter space: eigenvalue magnitude quantifies how strongly perturbations along each eigennetwork direction change the dynamical states of the fitted PMEM around the group reference model. Here, “stiff” and “sloppy” refer to model sensitivity with respect to parameter variations, as captured by the FIM, rather than to the strength of association with cognitive performance.

The parameter space, together with the eigennetworks as orthogonal directions, forms a sensitivity-based coordinate system for brain-behavior studies across individuals (Fig. 2F, G). To link sensitivity geometry to behavior, we computed each participant’s parameter deviation vector $\overrightarrow{\Delta\theta_q} = \overrightarrow{\theta_q} - \overrightarrow{\theta_g}$ relative to the group reference $\overrightarrow{\theta_g}$ and optimized a linear combination of eigennetworks to identify a single behavior-sensitive axis $\overrightarrow{v_{\mathrm{op}}}$ for a given task under three-fold cross-validation (Fig. 2F, right, see Method). In intuitive terms, this procedure searches for a “behavior-aligned direction” in sensitivity space: a weighted combination of eigennetworks along which participants’ parameter deviations become optimally ordered by task performance (Fig. 2G).We refer to $\overrightarrow{v_{\mathrm{op}}}$ as the **task-processing network**: a coordinated pattern of excitability and coupling weights along which individual deviations from the group reference best align with task performance. For each parameter, its absolute component magnitude $|v_{\mathrm{op},k}|$ quantifies its contribution to this task-processing network (for both $h$ and $J$ parameters), while its sign indicates whether better performance is associated with increased (positive) or decreased (negative) excitability or coupling. The scalar projection of $\overrightarrow{\Delta\theta}$ onto $\overrightarrow{v_{\mathrm{op}}}$ defines the *Behavior Aligned Individual Difference* (Behavior-AID), quantifying each individual’s deviation in brain parameters along this task-processing network axis relative to the group reference (Fig. 2F, left; conceptualized in Fig. 2G). A larger positive Behavior-AID indicates stronger alignment with the signed task-processing network pattern, rather than reflecting a uniform shift in any single global network property. In this way, SNM identifies a shared processing network $\overrightarrow{v_{\mathrm{op}}}$ and captures individual difference in task performance along this network via Behavior-AID.

Unless otherwise stated, we first focused on HCP working-memory (WM, n-back) performance, where “overall WM accuracy” denotes accuracy aggregated across 0-back and 2-back blocks, and “2-back accuracy” denotes accuracy within 2-back blocks only. We later apply SNM to the HCP language task to assess the generalizability.

Unlike MVPA and CPM, SNM summarizes coordinated variation in excitability and coupling in a single interpretable axis, the Behavior-AID, derived from a behavior-aligned combination of sensitivity-ranked eigennetworks (Fig. 2H), providing much greater predictability than MVPA and CPM. This framing sets up the central question for the following analyses: whether the neural parameter directions that matter most for brain activity and behavior are the same ones that account for the largest inter-individual variance in brain networks.

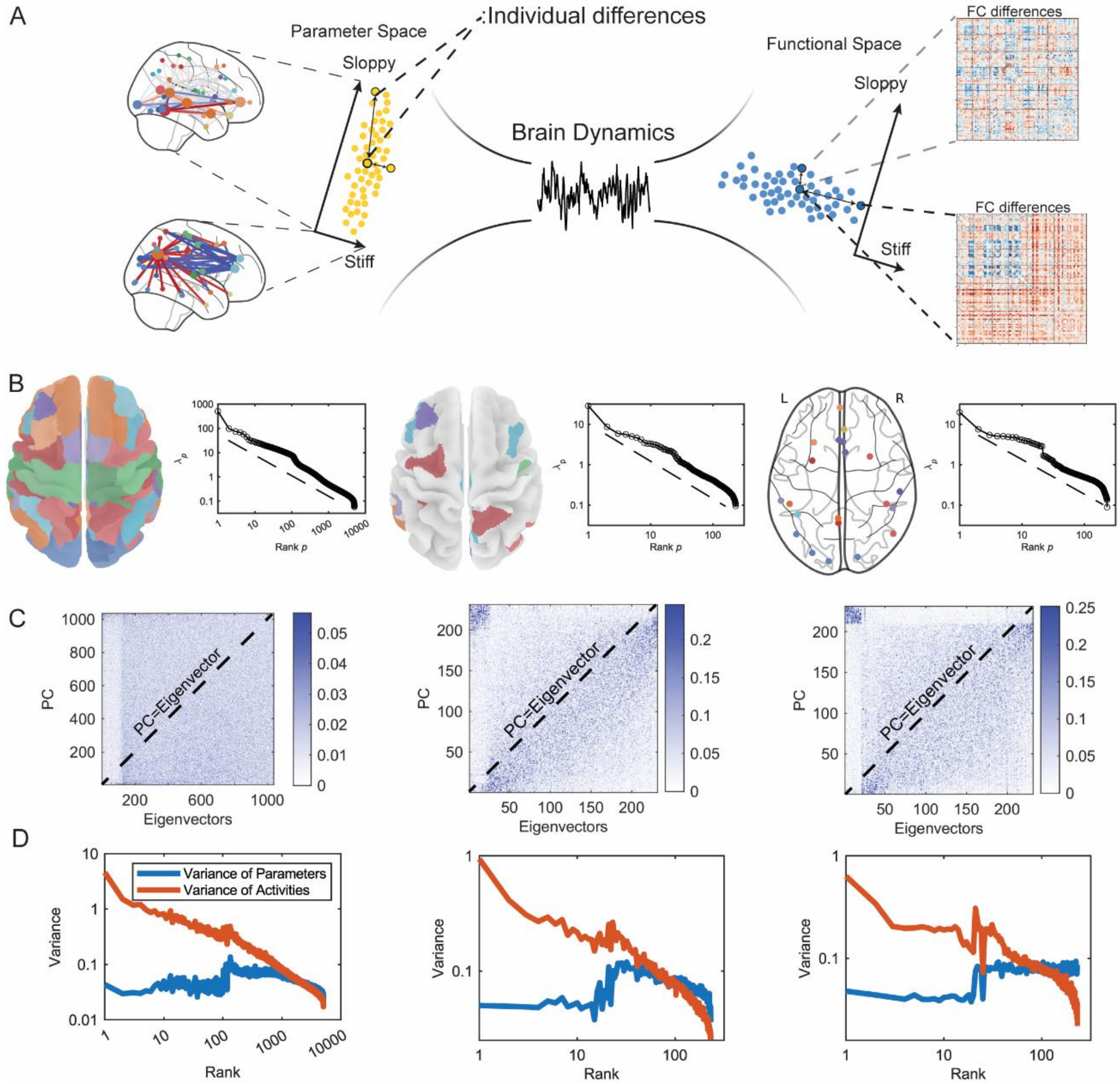

**Figure 3. Evidence for stiff-sloppy properties of brain: mismatch of parameter sensitivity and individual-level variance. (A)** Schematic illustration of stiff–sloppy geometry across brain network parameters and functional state space. In parameter space (yellow cloud), the major axis with large parameter variance aligns with a sloppy direction and the short axis aligns with a stiff direction; example eigennetwork connectivity patterns are shown alongside the axes. In the functional state space (blue cloud), the major axis aligns with the stiff direction, indicating amplified functional consequences along stiff dimensions even though the parameter variation is smaller. Example participant pairs separated primarily along the sloppy versus stiff axis are highlighted; for each pair, the functional connectivity (FC) difference is shown as a matrix, with stiff-axis differences exhibiting more systematic and modular structure**. (B)** Eigenvalue spectra of the group-level FIM across different system scales: left, cortex-only whole brain 100-region parcellation; sensitivity-guided reduced 21-region system selected by the largest sensitivity weights derived from $\overrightarrow{v_{\mathrm{op}}}$ ; right, activation-derived 21-ROI system assembled from literature-based working-memory task-positive and task-negative/DMN coordinates (see Methods). **(C)** Cosine-similarity matrices between PCA directions (rows) and eigennetworks (columns). The x-axis shows eigennetworks ranked by descending eigenvalue (rank 1 = largest eigenvalue = stiffest eigennetwork); the y-axis shows PCs ranked by descending loading magnitude (PC1=highest loading). The dashed diagonal line marks the locus where PCs and eigennetworks would coincide; the observed partial inversion indicates that stiff eigen-networks align more closely with lower-loading PCs rather than high-loading PC1/PC2. **(D)** Log–log plots of variance across individuals in

parameter-deviation projections (Variance of Parameters, blue lines) and FC-deviation projections (Variance of Activities, red lines) along the eigennetworks as a function of eigennetwork rank. Left to right: cortex-only whole brain, sensitivity-guided reduced 21-region system, and activation-derived 21-ROI systems.

## Behaviorally relevant directions do not coincide with large principal components of inter-individual variation

We next addressed this question by comparing the sensitivity-ranked eigennetworks with variance-based descriptions of inter-individual differences. In parameter space, we described inter-individual differences using subject-specific PMEM parameter deviations $\overrightarrow{\Delta\theta_q} = \overrightarrow{\theta_q} - \overrightarrow{\theta_g}$ relative to the group PMEM. In the full cortex-wide system, principal components analysis (PCA) of $\Delta\vec{\theta}$ showed that the principal components (PCs) capturing the largest inter-individual variation did not align with the eigennetworks with the strongest effects on brain dynamics; instead, the stiffest eigennetworks aligned more closely with lower-variance PCs (Fig. 3C). Thus, the neural directions along which individuals differ most were not the same as those with the greatest impact on model-predicted brain activity.

We observed the same mismatch when comparing how inter-individual variation is distributed across the eigennetwork hierarchy in parameter and functional space (Fig. 3A,D). In parameter space, we analyzed subject-specific PMEM parameter deviations relative to the group reference; in functional space, we analyzed subject-specific FC deviations relative to the group-averaged FC. Because these are distinct representational spaces, we did not compare their coordinates directly. Instead, we compared how inter-individual variance is distributed when each type of deviation is projected onto the same eigennetwork directions derived from the parameter-space FIM (Fig. 2G). Projecting individual parameter deviations and FC deviations onto eigennetworks revealed complementary variance loads: variance in parameter-space projections increased toward sloppier ranks, whereas variance in functional-space projections decreased toward sloppier ranks (Fig. 3D, left). In other words, larger inter-individual differences in parameters tended to lie in directions with weaker consequences for brain activity, whereas directions with stronger effects on brain activity carried less inter-individual variance. This dissociation directly supports the view that inter-individual variance and behavioral relevance are not equivalent.

Beyond this variance mismatch, the eigennetwork hierarchy itself also showed a clear spatial progression. Stiffer eigennetworks tended to involve broader, system-level patterns, whereas sloppier eigennetworks became increasingly localized and sparse, often resembling small motifs involving only a few regions and connections (Fig. 2F). Although the FIM yielded a very large number of eigennetworks ($M$=5,050 when considering parameters in the 100-region whole-brain model), the behavior-optimized combination (see Method) recruited only a small subset: on average ($\sim 165$) eigennetworks ($s.d. \sim 32$) carried non-negligible weights, corresponding to only $\sim 3.2\%$ of all eigennetworks across the sampling realizations (Supplementary Fig. S3). The optimization is implemented by recruiting increasingly sloppier modes only while they improved held-out prediction and stopped once additional modes no longer yielded meaningful gains (see Methods). This concentration of behavioral impact to a “small active subspace” already suggested that the dimensions most relevant to cognition do not coincide with the dominant structure of inter-individual variation.

We then asked whether this pattern depended on system scale or ROI definition. In the full cortex-only whole-brain system, the group-level FIM eigenspectrum showed a heavy-tailed, approximately power-law-like form (Fig. 3B, left). We repeated this analysis in two reduced systems: a sensitivity-guided 21-region subsystem derived from the whole-brain task-processing network (Fig. 3B, middle), and an activation-derived 21-ROI subsystem (Fig. 3B, right) defined from literature-based working-memory activation and DMN deactivation coordinates (detailed in Methods). Despite strong dimensional reduction and different region-selection criteria, both reduced systems retained similar heavy-tailed spectra, indicating that stiff-sloppy property is not an artifact of system size or a particular ROI definition. Moreover, the dissociation between parameter variance and functional consequences across eigennetwork ranks was preserved across both reduced systems.

Together, these analyses revealed a consistent mismatch between where individuals differ most and where changes have the largest consequences for brain activity. They also provided the rationale for the next step of the analysis: constructing a task-processing network from high-impact eigennetworks rather than from the dominant axes of inter-individual variation.

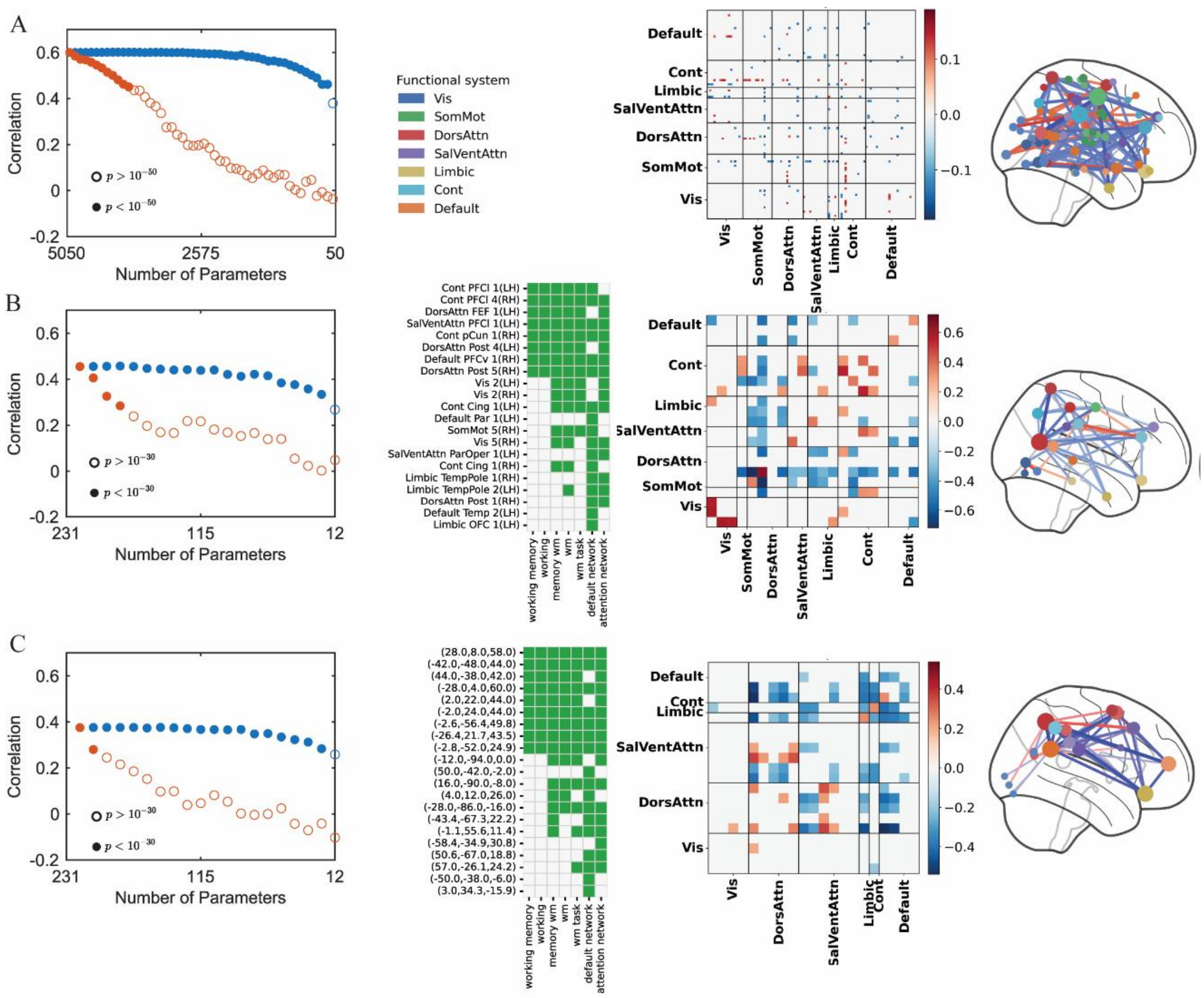

**Figure 4 | Sensitivity-based sparsification reveals a sparse, topologically coherent task-processing network. (A)** Cortex-only whole-brain (Schaefer-100) results. Sequential parameter removal based on sensitivity magnitude ($|v_{\mathrm{op,ij}}|$): parameters are removed either from lowest-to-highest sensitivity (blue) or highest-to-lowest sensitivity (red), and the association between Behavior-AID and task performance is evaluated as a function of the number of retained parameters (scatter plot). The corresponding sparsified sensitivity matrix (2%, color-bar denoting signed sensitivity) and brain-surface rendering visualize the sparse

and topologically coherent network structure retained after removing 98% low-sensitivity parameters (edges are coded by sign and magnitude; node size reflects degree (total sensitivity), colors denote functional systems). **(B)** Sensitivity-guided reduced top-21 system. Same removal analysis as in (A), along with the sparsified sensitivity matrix (20%) and brain-surface rendering. A binarized term-by-region matrix summarizes Neurosynth meta-analytic associations: green cells indicate positive associations between regions and functional terms, whereas white cells indicate no such association. **(C)** Activation-derived reduced system of 21 ROIs. Same analyses as in (B): sequential sensitivity-based removal, sparsified sensitivity matrix, Neurosynth term-by-region association matrix (green = positive association; white = no association), and brain-surface rendering of the sparsified network (20%).

## Removing low-sensitivity variance preserves behavioral explanation

Having identified a task-processing network that captures the coordinated parameter differences most relevant to behavior, we next asked how much of the inter-individual variation in this parameter space is actually needed to explain performance. If much neural variation is behaviorally weak, then removing low-sensitivity parameters should have little effect on the association between Behavior-AID and task performance, whereas removing high-sensitivity parameters should rapidly degrade it. To test this idea,, we performed sequential removal analyses in three systems (Fig. 4). In each system, we progressively removed parameters either (i) from lowest-to-highest parameter sensitivity or (ii) from highest-to-lowest parameter sensitivity, and recomputed Behavior-AID and quantified the association between Behavior-AID and performance as a function of the number of retained parameters.

In the cortex-only whole-brain system, removing the low-sensitivity parameters preserved the Behavior-AID–performance relationship across a wide range, whereas removing the high-sensitivity parameters led to a rapid loss of the association (Fig. 4A). Notably, retaining only a small fraction of the most sensitive parameters (e.g., 2%) preserved a strong association. A supporting network-integrity analysis showed that the size of the giant component declined along a trajectory closely matching the slow-degradation regime under removal of low-sensitivity parameters (Supplementary Fig. S3B). Together, these results indicate that the Behavior-AID–performance association remains preserved while the sensitivity-defined network retains a coordinated large-scale structure, unless this structure is substantially disrupted.

The corresponding sparsified sensitivity structure formed a distributed, but topologically coherent subnetwork in both matrix and surface renderings (Fig. 4A), that is, the retained high-sensitivity edges preferentially assemble into a connected backbone with clustered neighborhoods and consistent long-range bridges, rather than scattered isolated links. The high-sensitivity core prominently involved control/attention hubs such as Control PFCl 1 (LH), Control PFCl 4 (RH), DorsAttn FEF 1 (LH), Control pCun 1 (RH), and DorsAttn Post 4/5, consistent with frontoparietal–dorsal-attention architectures repeatedly implicated in working-memory control (Cole et al. 2013, Duncan 2010, Rottschy et al. 2012). To quantify the decline in behavioral explanation during sequential removal, we defined a decline metric, $D(n) = 1 - R(n)/R(0)$, where $R(n)$ denotes the Behavior-AID–performance correlation after removing $n$% of parameters (Supplementary Fig. S3A). In the cortex-only whole-brain system, $D(n)$ increased only gradually when low-sensitivity parameters were removed, reaching a 20% drop in performance association only after removing approximately 95% of parameters, whereas the same 20% drop was reached already after removing only approximately 20% of high-sensitivity parameters.

We observed the same sensitivity-dependent degradation pattern in the sensitivity-guided reduced system (Fig. 4B) and the activation-derived system (Fig. 4C). Functional annotation using

Neurosynth term–region associations (Yarkoni et al. 2011) further indicated that sensitivity-preserving sparsification retained coherent cognitive profiles across reduced systems (Fig. 4B–C). Compared with both reduced systems, the cortex-only whole-brain system showed a stronger brain–behavior association (Fig. 4, correlation ~ 0.6 vs ~0.4), suggesting that although a sparse and topologically coherent high-sensitivity core captures most behavior-aligned signal, additional intermediate-sensitivity elements in the full system still contribute meaningful predictive information. This sparsification result helps explain why a relatively small number of retained parameters (and the corresponding subnetwork) can still account for behavior. The high-sensitivity subset is not merely a small set of individually strong features, but preserves a coordinated pattern of parameter differences whose system-level effects and behavioral relationships are retained in the task-processing network. By contrast, much of the remaining inter-individual variation can be removed with little consequence for behavioral explanation. Thus, the task-processing network distinguishes behaviorally consequential variance from variance that is substantial across individuals but weak in its effects on brain activity and performance.

We obtained the same qualitative pattern when the optimization target was changed. In working memory, separate optimizations for overall accuracy, 0-back accuracy and 2-back accuracy all reproduced the same asymmetry: removing low-sensitivity parameters preserved the AID–behavior association over a broad range, whereas removing high-sensitivity parameters caused rapid degradation (Supplementary Fig. S4A–B). The same logic generalized to the HCP language task, where overall, story, and math accuracy all showed the same sensitivity-guided removal asymmetry (Supplementary Fig. S5A–C). Thus, the concentration of behavioral leverage in a sparse high-sensitivity subnetwork generalized across tasks, and importantly, the detailed regional composition of the resulting task-processing networks is task-specific, demonstrating precision association from un-biased whole-brain dynamics patterns.

## Edge-wise feature selection misses coordinated behaviour-relevant variation and yields unstable feature maps

We next asked why variance-based or edge-wise summaries can retain substantial neural variation yet still fail to recover the network structure that generate behavior. In a coupled brain system, an individual connection can show only weak marginal association with behavior when treated independently, while still belonging to a coordinated parameter combination with strong consequences for brain activity and performance. We therefore compared SNM-derived parameter sensitivity with the edge-wise FC–behavior associations used in CPM.

We related the correlation between WM performance and the $FC_{ij}$ in CPM to its parameter sensitivity rank derived from the task-processing network $\overrightarrow{v_{op}}$ by SNM. Many edges passing the CPM significance threshold occupied low-to-intermediate sensitivity ranks (blue dots in Fig. 5A), whereas numerous high-sensitivity edges did not pass edge-wise significance when treated as independent predictors (red dots in Fig. 5A). Thus, edge-wise statistical association is not equivalent to behavioral leverage revealed by SNM under coupled network dynamics. In particular, many high-sensitivity edges were missed by CPM even though they formed part of the coordinated network pattern identified by the task-processing network. This result indicates that treating connections as independent indicators can preserve variance that is statistically associated with behavior while missing the network structure that is most behaviorally consequential.

We then quantified how this representational difference affected predictive performance. We found that CPM predictability changed as the $p$-value threshold for edge-task association became more stringent (from $p$=0.1 to $10^{-6}$), dramatically reducing feature counts. CPM predictability remained relatively low and did not systematically improve with increasing thresholds. Crucially, when the $p$-threshold decreased below 0.001 (feature number < 741), CPM predictability suddenly dropped (Fig. 5B). In contrast, SNM predictability remained high over a wide range when low-sensitive parameters were removed and maintained a significantly high predictability even with only a small number of features (303; 6% features); it dropped primarily when high-sensitive parameters were removed (Fig. 5B). Sensitivity-aware CPM controls confirmed this pattern. Adding a small number (1.6%) of SNM-identified high-sensitivity edges (red dots in Fig. 5A) to CPM as independent indicators significantly improved CPM predictability (Enhanced-CPM; Fig. 5C). Using only 2.9% high-sensitivity edges (red and green dots) as independent CPM predictors achieved similar performance as standard CPM (p<0.01, with 26.6% of edges) despite using far fewer edges (Sensitive-only; Fig. 5C). However, the highest predictability was obtained when this small set of high-sensitivity edges were combined in the SNM-coordinated manner defined by the task-processing network $\overrightarrow{v_{\mathrm{op}}}$ (Sparse SNM; Fig. 5C). These controls show that performance depends not only on which network elements are retained, but also on whether they are evaluated and combined as parts of a coordinated system.

If edge-wise feature selection misses coordinated behavioral-relevant variation, then the resulting feature maps should also be less stable under resampling. We therefore compared the stability of SNM-derived task-processing networks with CPM-selected feature sets under matched repeated three-fold cross-validation procedures. We estimated $\overrightarrow{v_{\mathrm{op}}}$ across repeated three-fold cross-validation runs and computed pairwise cosine similarity between the resulting task-processing networks. For a fair comparison, CPM feature sets were evaluated under matched resampling procedures and the same held-out performance metric. The SNM-derived task-processing networks showed markedly higher cross-run similarity than CPM feature sets (Fig. 6A). This stability difference was also evident in matrix and brain surface visualizations: $\overrightarrow{v_{\mathrm{op}}}$ retained consistent structure across runs, whereas CPM-selected edge sets varied substantially with resampling trials (Fig. 6B). Out-of-sample predictability, evaluated as test-fold correlation between predicted and experimentally observed WM scores across repeated cross-validation, was much higher for SNM than for CPM (Fig. 6C).

Together, these results show that instability in conventional feature maps is not only a sampling problem, but also a representational one. When connections are treated as independent features, feature selection can emphasize marginally associated edges while missing the coordinated network variation that supports behavior. By modeling individual differences as coordinated deviations along a behaviorally relevant network axis, SNM provides a more stable and generalizable representation of brain–behavior association.

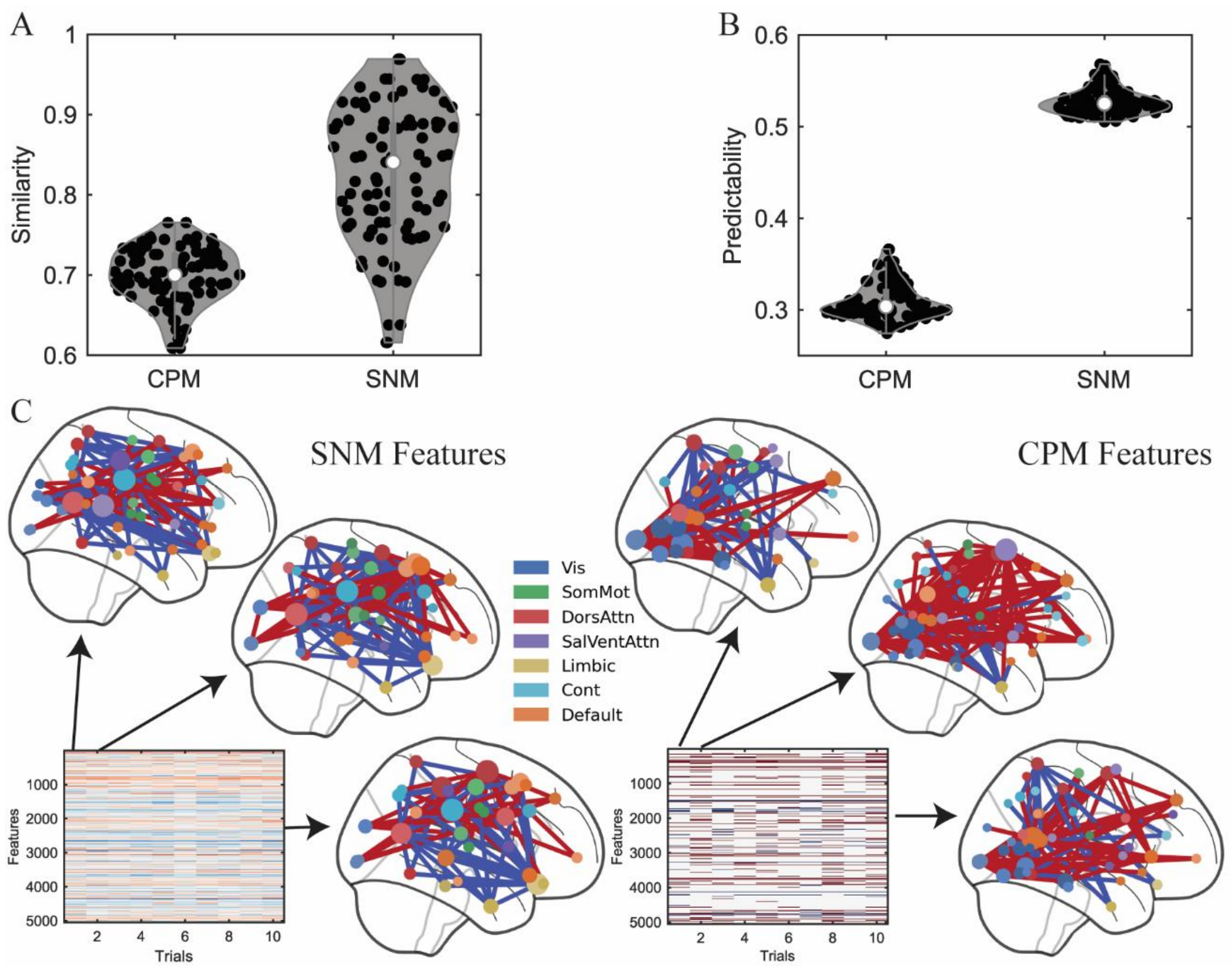

**Figure 6 | Stability of task-processing networks and comparison of predictive performance between SNM and CPM. (A)** Distributions of pairwise cosine similarities of the processing networks across independent three-fold cross-validation runs (10 runs, 90 pairwise similarities): left: similarities among SNM-derived WM processing networks $\overrightarrow{v_{\mathrm{op}}}$ ; right: similarities among CPM-selected significant-connection sets obtained under the same cross-validation scheme. Each point represents one pair of runs. **(B)** Distributions of test-set performance *predictability* for SNM and CPM across 100 independent three-fold cross-validations. For SNM, $\overrightarrow{v_{\mathrm{op}}}$ is identified in training folds, Behavior-AID is computed for all subjects and used to predict task scores by training a linear regression model; predictability is defined as the correlation between predicted and experimentally observed scores in the held-out fold. CPM predictability is evaluated similarly using CPM-selected features. **(C)** Matrix representations of $\overrightarrow{v_{\mathrm{op}}}$ (SNM) and CPM selected-connection sets across repeated runs (columns = runs; rows = connections). Brain-surface renderings of binarized SNM and CPM networks for visualization (node size reflects degree; links correspond to the selected-connection sets. Red/blue color: positive/negative sensitivity in SNM or positive/negative edge weight-WM score correlation in CPM.

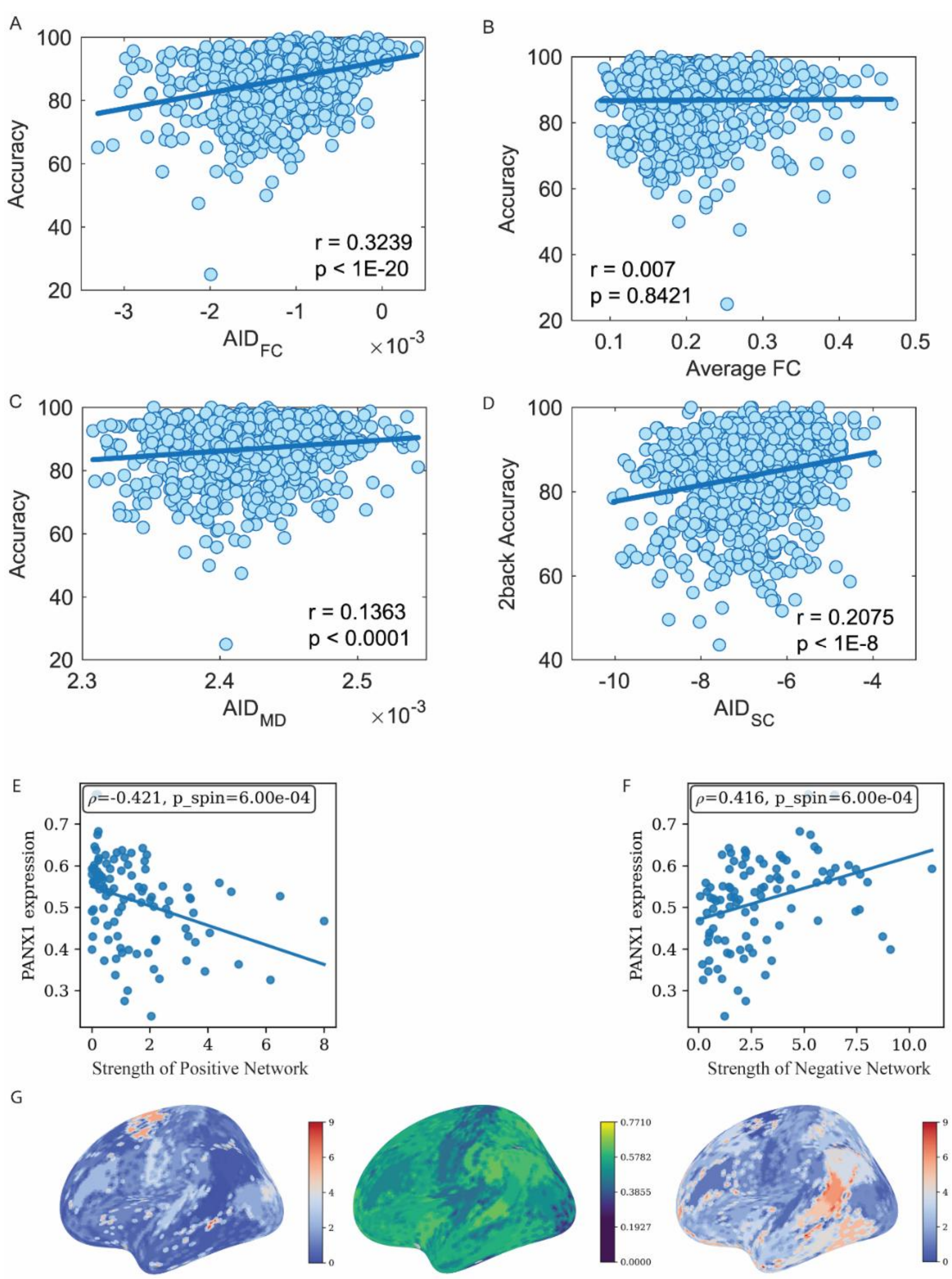

**Figure 7 | Multimodal and transcriptomic correlates of the SNM-revealed task processing network $\overrightarrow{v_{\mathrm{op}}}$.** **(A)** Sensitivity-weighted functional connectivity alignment denotes variation of individual FC along $\overrightarrow{v_{\mathrm{op}}}$. For each participant, element-wise multiplication of $\overrightarrow{v_{\mathrm{op}}}$ with the participant's FC yields a weighted FC matrix; summing weighted FC defines $AID_{FC}$, which correlates with overall WM accuracy (Pearson *r*=0.3239, $p < 10^{-20}$). **(B)** Unweighted FC summaries do not explain performance. Direct summation of FC shows no association with task performance (Pearson *r*=0.007, *p*=0.8421). **(C)** Sensitivity-weighted regional microstructure alignment denotes variation of individual regional microstructure along $\overrightarrow{v_{\mathrm{op}}}$. Using the regional ($h$) component of $\overrightarrow{v_{\mathrm{op}}}$, element-wise multiplication with each subject's regional mean diffusivity (MD) vector and summation across regions yield $AID_{MD}$, which correlates with overall WM accuracy (Pearson *r*=0.1363, *p*<0.0001). **(D)** Sensitivity-weighted structural connectivity (SC) alignment denotes

variation of individual SC along $\overrightarrow{v_{\text{op}}}$ optimized for 2-back task accuracy. After removing the diagonal of the SC matrix, element-wise multiplication of $\overrightarrow{v_{\text{op}}}$ with SC and summation yield $AID_{SC}$, which correlates with 2-back accuracy (Pearson $r$=0.2075, $p < 10^{-8}$). **(E)** Scatter plot of regional strength in the positive subnetwork of $\overrightarrow{v_{\text{op}}}$ optimized for 2-back task accuracy versus regional PANX1 gene expression (Spearman ρ; significance assessed with spatial spin tests). **(F)** Same as (E), but for the negative subnetwork; the PANX1 association is of opposite sign. **(G)** Cortical surface maps showing sensitivity positive-subnetwork strength (left), PANX1 expression (middle), and sensitivity negative-subnetwork strength (right).

## The task-processing network provides a common coordinate system across modalities

Having identified a task-processing network that distinguishes behaviorally consequential from behaviorally weak parameter variation, we next asked whether the same network could also organize inter-individual differences across multiple modalities. If the task-processing network captures how coordinated differences in excitability and coupling shape behavior, then functional, structural, and molecular variation aligned with this network should be more informative than unweighted global summaries.

We first tested this idea in functional connectivity (FC). We computed a sensitivity-weighted alignment score (projection along $\overrightarrow{v_{\text{op}}}$, Fig. 2G) by element-wise multiplying each participant's FC matrix with $\overrightarrow{v_{\text{op}}}$ and summing the weighted entries, yielding $\text{AID}_{\text{FC}}$. $\text{AID}_{\text{FC}}$ correlated significantly with WM accuracy (Pearson ($r$=0.3239, $p < 10^{-20}$; Fig. 7A). In contrast, unweighted global FC summaries showed no association with performance (Pearson $r$=0.007, p=0.8421; Fig. 7B). Similar null results were obtained for direct unweighted summaries within sensitivity positive/negative subnetworks (Supplementary Fig. S6), but sensitivity-weighted FC separately for sensitivity positive and negative subnetworks show significant correlation with WM accuracy (Supplementary Fig. S7). These results indicate that functional variation becomes behaviorally informative not by its magnitude alone, but by how it is weighted by the task-processing network. Because FC is estimated from the full task (0-back and 2-back) time series and reflects the network's moment-to-moment coordination across both load conditions, we report FC alignment primarily with the task-processing network optimized for overall WM accuracy (0-back + 2-back), consistent with the main WM analyses above.

We then asked whether structural variation aligns with the same behaviorally relevant coordinate system. Because structural and microstructural measures are more trait-like and are expected to relate more strongly to the working-memory-specific component of performance, we used the 2-back-optimized task-processing network. Using its regional component in $\overrightarrow{v_{\text{op}}}$, we computed $\text{AID}_{\text{MD}}$ by weighting each participant's regional mean diffusivity (MD, see Method) as a measure of microstructure of cortical regions and summing across regions. $\text{AID}_{\text{MD}}$ correlated significantly with WM 2-back accuracy (Pearson $r$=0.1363, $p$<0.0001; Fig. 7C). For structural connectivity, we computed $\text{AID}_{\text{SC}}$ by weighting each participant's SC matrix (diagonal removed) by $\overrightarrow{v_{\text{op}}}$ optimized for 2-back task accuracy and summing, capturing coordinated SC variation aligned with the task-processing network of 2-back WM task. $\text{AID}_{\text{SC}}$ correlated with 2-back accuracy (Pearson $r$=0.2075, $p < 10^{-8}$; Fig. 7D). Thus, structural and microstructural variation became more behaviorally interpretable when expressed in the same sensitivity-weighted coordinate system that was derived from the task-processing network.

We next examined whether this coordinate system also aligns with cortical molecular organization. We split $\overrightarrow{v_{\text{op}}}$ into sensitivity positive and negative subnetworks and computed regional strength maps (summing sensitivity of parameters within and linking to each region). We then related these maps to cortical gene-expression profiles (abagen-processed AHBA data) (Markello et al. 2021) and assessed significance using spatially constrained spin tests (see Methods). The most robust association ($\overrightarrow{v_{\mathrm{op}}}$ optimized for 2-back task accuracy is shown in Fig. 7E-G; see Supplementary Fig. S8 for other conditions) was obtained with PANX1 expression correlating negatively with sensitivity strength in the positive subnetwork and positively with strength in the negative subnetwork (Fig. 7E–G). PANX1 encodes pannexin-1, a large-pore membrane channel that supports ATP release and purinergic/adenosine signaling, and has been implicated in regulating synaptic transmission and plasticity and the excitatory–inhibitory balance (Illanes-Gonzalez et al. 2025, Prochnow et al. 2012). In this context, the opposite-signed PANX1 associations for the positive and negative subnetworks are consistent with a molecular gradient that could differentially couple to network components supporting network integration versus segregation, potentially via regionally varying constraints on excitability and synaptic plasticity. Cortical maps confirmed that PANX1 expression follows the spatial topography of the sensitivity-defined integration–segregation components of the WM task-processing network (Fig. 7G).

## Discussion

A central issue highlighted by our results is the mismatch between inter-individual neural variation and behavioral relevance. By separating two distinct properties of brain differences across individuals, how much network parameter components vary and how strongly coordinated parameter changes alter network dynamics in behavior-aligned ways, SNM provides a principled account of this dissociation. Across working-memory and language tasks, we found that behaviorally relevant individual differences were concentrated in a limited set of coordinated high-sensitivity eigennetworks, whereas much larger inter-individual variation was distributed across low-sensitivity dimensions. This mismatch between inter-individual variation and behavioral relevance provides a unifying account of the main findings in our study. It explains why large neural differences can have little consequence for behavior, why a relatively small set of coordinated parameters can retain strong explanatory power, and why a task-processing network derived from sensitivity-ranked eigennetworks provides a stable precision representation of brain–behavior association with high predictability.

This dissociation also helps clarify a broader difficulty in brain–behavior association research. Effect sizes for complex cognition and mental health phenotypes are typically small, and stable brain–behavior associations often require sample sizes in the thousands (Marek et al. 2022). At the same time, the behavioral tasks that produce robust experimental effects can be paradoxically poor at supporting reliable between-subject inference, since the very success of an experimental manipulation often implies constrained between-person variance in the targeted process (Hedge et al. 2018). Our results suggest that one reason for this difficulty is representational rather than purely statistical: If behaviorally consequential variation is concentrated in coordinated dimensions that are not the largest in variance across individuals, then approaches that prioritize variance, edge-wise association, or covariance structure may systematically emphasize neural differences that are measurable yet behaviorally weak (Gutenkunst et al. 2007). Our results show that sensitivity-based geometry explicitly captures this dissociation, identifying a small number of coordinated parameter combinations that dominate output sensitivity, reflecting stiff-sloppy property (Gutenkunst et al. 2007, Transtrum et al.). Prior work shows that this principle is expressed across multiple levels of neural organization: spontaneous neuronal networks exhibit marked sloppiness in their collective dynamics (Panas et al. 2015), cortical state transitions and stimulus-evoked responses evolve preferentially along different stiff versus sloppy parameter dimensions (Ponce-Alvarez et al. 2020), and whole-brain wakefulness-to-anesthesia transitions can be captured by a restricted subset of sensitive macroscopic couplings (Ponce-Alvarez et al. 2022). Our results show that large-scale brain dynamics exhibit stiff–sloppy organization: eigennetworks that most strongly reshape the model-predicted state distribution, and align with cognitive differences across individuals, need not correspond to the directions of greatest inter-individual variance (Fig. 2). This extends sloppiness from spontaneous activity and brain-state transitions to inter-individual cognition, suggesting that the challenge of brain–behavior association is not simply that effect sizes are small, but that behaviorally relevant neural variation may be masked by larger yet less consequential forms of inter-individual variation.

A key implication of our results is that individual differences may be better understood as coordinated deviations from a shared group reference than as diffuse regional or edge-wise fluctuations. This perspective also helps explain why edge-wise or variance-based feature summaries can be unstable and difficult to interpret in correlated, high-dimensional brain data. In such settings, fitted weights and selected features often reflect how predictive information is distributed across correlated variables rather than which neural components are mechanistically most consequential (Haufe et al. 2014, Mihalik et al. 2022, Varoquaux 2018). Even more pointedly, signatures that are excellent for identifying individuals (e.g., connectome "fingerprints" (Finn et al. 2015)) can rely on functional systems distinct from those supporting behavioral prediction—underscoring that discriminability, variance, and behavioral relevance need not coincide (Mantwill et al. 2022). Our comparison with CPM makes this point explicit: many edges with high behavioral impact were not individually significant when treated as independent predictors, whereas many selected edges occupied low-to-intermediate sensitivity ranks, yielding feature sets that were less stable across resampling and less predictive out of sample. By contrast, SNM identifies behaviorally relevant variation as a coordinated network pattern rather than as a collection of independent edges. Consistent with broader evidence that neural systems exploit structured low-dimensional control subspaces (Cunningham et al. 2014), the task-processing network defines a behaviorally relevant axis in interpretable parameter space, and Behavior-AID quantifies where each individual lies along that axis. This representation is low-dimensional, but unlike statistical compression alone, it is organized by inferred consequences for brain dynamics and behavior, providing a compact description of which coupled changes in excitability and interregional coupling are most consequential for task performance. The explanatory power of a relatively small number of high-sensitivity parameters therefore lies not only in the parameters themselves, but in the structured way they act together on the whole system.

The same logic extends beyond functional task dynamics. Once individual variation is weighted according to its behavioral consequences, the task-processing network also provides a common coordinate system across modalities. Sensitivity-weighted projections reveal behaviorally meaningful variation in functional and structural connectomes and microstructure that is not apparent in global summaries, consistent with the idea that macroscale dynamics are constrained by organized anatomical structures (Alexander-Bloch et al. 2013, Paquola et al. 2019, Seidlitz et al. 2018). Extending to molecular organization, the association of WM network with PANX1 expression suggests that transcriptomic gradients can bias regional participation in integrative versus segregative coupling patterns, plausibly via ATP release and purinergic signaling mechanisms that modulate excitability and synaptic efficacy (Illanes-Gonzalez et al. 2025, Prochnow et al. 2012). This multimodal convergence suggests that the mismatch between inter-individual variation and behavioral relevance is not confined to a single measurement scale. Instead, behaviorally relevant individual differences appear to be organized by coordinated network effects that span structure, dynamics and molecular architecture. Overall, these analyses show that the task-processing network is more than a functional predictor of behavior. Rather than treating inter-individual variation as a diffuse collection of modality-specific differences, it organizes that variation according to its behavioral relevance. In this sense, sensitivity-based weighting does not simply identify better features within each modality; it reveals a common organization of multi-scale individual variation around behaviorally consequential network effects. This common coordinate system relating functional connectivity, structural connectivity, cortical microstructure, and transcriptomic gradients to a shared axis of performance-relevant network change thus represents a new framework of structure-function-behavior relationship.

The regional composition of the high-sensitivity core broadly overlaps with established control- and attention-related systems implicated in working memory, e.g., control/attention hubs such as Control PFCl 1, Control PFCl 4, DorsAttn FEF, Control pCun 1, and DorsAttn Post 4/5 (Cole et al. 2013, Duncan 2010, Rottschy et al. 2012) .The results also bridge the gap between region-based views of functional specialization (Bassett et al. 2017, Kanwisher 2010) and network-level accounts of connectivity (Bassett et al. 2017, Lee et al. 2022, Leergaard et al. 2022, Stern 2022, Thiebaut de Schotten et al. 2022). Supplementary analyses in the language task indicate that the same sensitivity-based logic generalizes beyond one cognitive domain. Importantly, however, our results do not imply that only a few regions matter or that low-sensitivity variation is universally irrelevant. Rather, they indicate that behaviorally consequential variation is concentrated in a sparse but coordinated subnetwork embedded within a broader system. Intermediate-sensitivity elements can still contribute, but they do so relative to the network pattern in which they are embedded according to cognitive demands, not as independent features. For example, the task-processing network optimized to language-task overall accuracy was associated with a broader default–control–salience–attention backbone (Cai et al. 2024), whereas story accuracy placed greater weight on bilateral temporal, temporal-pole, and default-network territories (Ralph et al. 2017, Zhu et al. 2025), and math task condition more strongly emphasized dorsal-attention, posterior parietal, and control territories, consistent with their distinct cognitive demands (Ansari 2008, Nieder 2016, Vogel et al. 2021). This distinction is important for precision brain association studies because it shifts the focus from identifying isolated "important" regions or edges to identifying the coordinated network variations that carry behavioral consequence.

More broadly, SNM is readily applicable to other cognitive tasks, brain functions (e.g., social-emotional processing) and phenotypes (e.g., development and aging, and disorders) to reveal the underlying processing networks. It will be important to study how such stiff-sloppy organization is related to other pronounced features of the brain, such as critical states (Haimovici et al. 2013, Tagliazucchi et al. 2012) and hierarchal modular organization (Wang et al. 2019) and balance of segregation and integration (Wang et al. 2021) in supporting diverse cognitive abilities, as well as deeper biological mechanisms, such as excitation-inhibition balance (Liang et al. 2025, Yizhar et al. 2011), which is increasingly being linked to macroscale structure–function coupling, cortical maturation and cognitive ability across the human cortex (Fotiadis et al. 2023, Zhang et al. 2024).

One particularly promising direction is sensitivity-based neuromodulation. Stimulation effects are increasingly recognized as network-mediated and state-dependent, with substantial heterogeneity across individuals even under nominally identical protocols (Beynel et al. 2020, Bradley et al. 2022, Watanabe et al. 2025). Connectivity-informed targeting has already shown that clinical efficacy can depend on intrinsic connectivity between the stimulation site and downstream network nodes—demonstrating that "where you stimulate" should be defined in network terms rather than purely anatomical localization (Fox et al. 2012). Our framework sharpens this logic for cognitive interventions such as working-memory enhancement (Lee et al. 2022, Merk et al. 2025, Reimer et al. 2024): rather than targeting regions with strong activation or marginal associations, interventions should aim to perturb the system along stiff directions. This yields testable, individualized hypotheses: (i) baseline Behavior-AID might predict who benefits most from stimulation or training; (ii) effective protocols should preferentially move individuals along the task-processing axis; and (iii) stimulation locations (or multi-site strategies) should be chosen to maximize alignment with the stiff sensitivity pattern, potentially explaining why nominally similar working-memory TMS protocols can show variable effects (Hedge et al. 2018). More broadly, this framework may provide a theoretical basis for individualized, network-guided, and potentially multi-site neuromodulation strategies (Merk et al. 2025, Shin et al. 2024, Wei et al. 2025).

Several limitations motivate clear further steps. First, although SNM is based on a dynamical generative model and network parameters and their structural association point to the generative and mechanistic directions of brain-behavior relationship, explicit causal validation is essential. The strongest test of a sensitivity-defined dimension is whether experimentally induced perturbations, e.g., via brain stimulation, pharmacological intervention, or learning, aligned with the identified task-processing axis produce larger and more reliable behavioral changes than non-aligned perturbations of similar magnitude. Second, although the present results focus on working memory and language in healthy young adults, the framework makes specific predictions across domains and populations: different tasks and disorders should be characterized by distinct stiff subspaces, as well as by distinct cross-modality, cross-scale correspondences, for example among functional dynamics, structural connectivity and microstructure, and transcriptomic organization, and these predictions can be adjudicated in longitudinal and clinical datasets. Third, sensitivity estimation in very high-dimensional connectomes can be computationally demanding; however, the multiscale consistency (whole-brain vs. activation-defined subnetwork) observed here suggests that dominant behavior-relevant dimensions can be approximated in reduced representations, enabling practical deployment while preserving mechanistic interpretability. Together, these next steps will determine the causal validity, generality, and practical usefulness of sensitivity-defined dimensions for understanding brain–behavior relationships.

# Conclusion

## xxx

# Methods

**Details of fMRI data acquisition and preprocessing**

We utilized data from the HCP (Van Essen et al. 2013) containing structural MRI, diffusion tensor imaging (DTI), resting and task state fMRI, and behavioral measures on multiple cognitive tasks for 1,200 healthy young adults (Van Essen et al. 2013) database (http://www.humanconnectome.org). Neuroimaging data had been acquired using a Siemens Skyra 3T scanner and preprocessed in accordance with standard HCP protocols. The WM task used in the HCP follows a block design with alternating 2-back working-memory and 0-back attention-control conditions. Each block began with a 2.5 s written cue indicating the task type (2-back or 0-back) and, in the case of the 0-back block, the pre-specified target stimulus for that block. Stimuli from four visual categories (faces, places, tools, and body parts) were presented in separate blocks. Each stimulus was displayed for 2 s, followed by a 500 ms inter-stimulus interval. During 2-back trials, participants were instructed to respond whenever the current stimulus matched the one presented two trials earlier, thus requiring continuous stimulus monitoring, updating and maintenance of the stimuli and their presentation order in working memory. In contrast, 0-back trials involved responding to a pre-identified stimulus whenever it appeared, requiring continuous attention (alertness). The 0-back task involves the same stimulus and response processes as the 2-back task and may therefore serve as a control condition devoid of updating and order memory. For the mixed-condition analyses aiming at analyzing task-positive and task-negative network activities, we concatenated the blocks of 0- and 2-back tasks into a single time series; for the condition-specific analyses shown in Fig. 7, blocks of each task were modeled independently.

For the HCP language task, we analyzed the standard HCP auditory story–math paradigm. Each of the two runs interleaves four story blocks and four math blocks. Story blocks present brief auditory stories (typically 5–9 sentences, adapted from Aesop's fables), followed by a two-alternative forced-choice comprehension question about the story theme. Math blocks are also auditory and require participants to solve a series of addition/subtraction operations, followed by a two-alternative forced-choice response. Block lengths vary (average ~30 s), and the math blocks were designed to approximately match the story-block durations; additional math trials may be appended to complete the run (Barch et al. 2013).

To analyze HCP task-based fMRI data, we utilized the FEAT analysis tool after applying the HCP minimal preprocessing pipelines (Glasser et al. 2013). For scan-level analyses, we based our data processing approach on the provided level1.fsf template and customized it for our purposes. Spatial smoothing was applied with a full-width at half maximum (FWHM) of 4 mm, and a high-pass filter cutoff of 90 s was used to remove low-frequency noise, as determined by design efficiency estimates calculated within FEAT. Motion correction was performed using MCFLIRT, ensuring precise alignment of images. Registration steps were excluded from the Level 1 analysis because the HCP preprocessing pipelines had already aligned the data to the MNI152 template. The final output consisted of fully filtered 4D fMRI data, which was used for subsequent analyses.

Subject inclusion was defined separately for each task. For WM analyses, we retained subjects who simultaneously had usable n-back accuracy records and structural data required for the multimodal analyses. For language analyses, we retained subjects who simultaneously had both story-accuracy and math-accuracy records. These behavioral/data-availability criteria were applied before the PMEM fitting-quality criteria described below; the final retained sample for each analysis was then further constrained by PMEM convergence thresholds.

**Diffusion MRI, structural connectivity, and NODDI processing**

Diffusion MRI analyses were based on the HCP S1200 minimally preprocessed diffusion data. The HCP diffusion pipeline includes across-run b0 intensity normalization, correction of susceptibility-induced distortions, eddy-current distortions, subject motion, and gradient nonlinearity, followed by registration of diffusion data to the T1-weighted structural image, transformation into 1.25-mm native structural space, and application of a final brain mask (Glasser et al., 2013; Sotiropoulos et al., 2013). Prior to downstream analyses, diffusion gradient tables underwent quality control to ensure correct correspondence with diffusion volumes, including shell verification, removal of corrupted entries where necessary, and normalization of gradient vectors.

NODDI-Based Microstructural Estimation

Microstructural properties were estimated using the NODDI framework (Zhang et al., 2012), implemented with the AMICO toolbox (Daducci et al., 2015). NODDI fitting was performed to derive voxel-wise maps of neurite density index (NDI), orientation dispersion index (ODI), and isotropic volume fraction (ISOVF), with the intrinsic parallel diffusivity fixed at a gray-matter-appropriate value for cortical estimation. The resulting metric maps were transformed to MNI152 standard space using FSL applywarp and subsequently projected to the 32k fs_LR cortical surface using ribbon-constrained volume-to-surface mapping in Connectome Workbench. Parcel-wise mean values were then extracted for each metric using the Schaefer-100 atlas.

Structural Connectivity

Structural connectivity matrices were derived from the same HCP S1200 diffusion data using an advanced tractography pipeline. Following minimal preprocessing, multi-tissue response functions for white matter, gray matter, and cerebrospinal fluid were estimated using the dhollander algorithm (Dhollander et al., 2019) in MRtrix3 (Tournier et al., 2019). Multi-shell multi-tissue constrained spherical deconvolution was then applied to estimate fiber orientation distributions. Whole-brain probabilistic tractography was performed using the iFOD2 algorithm with anatomically constrained tractography based on a five-tissue-type segmentation derived from the T1-weighted image, generating ten million streamlines. Streamline weights were subsequently refined using SIFT2 to improve the biological plausibility of reconstructed connections (Smith et al., 2015). The Schaefer-100 cortical parcellation was transformed into each subject's native diffusion space, and structural connectivity matrices were constructed using the sum of SIFT2 streamline weights between parcel pairs.

**ROI systems and reduced-system definitions**

We used three systems in parallel: (i) a cortex-only whole-brain system (Schaefer et al. 2018, Yeo et al. 2011) (Schaefer-100), (ii) a sensitivity-reduced 21-region system, and (iii) an activation-derived 21-ROI system.

The sensitivity-reduced system was defined post hoc from the SNM-identified task-processing network from the un-biased cortex-only whole-brain system by selecting the 21 regions with the largest regional sensitivity summaries (computed from the absolute weights in task processing network)).

The activation-derived 21-ROI system was defined to provide an external, literature-based comparison system rather than a sensitivity-based selection. We created spherical ROIs centered on these published coordinates (12-mm diameter spherical ROI, coordinates list in Supplementarty Table. 1) and extracted mean ROI time series for the reduced-system analyses. Importantly, the coordinates were assembled from two sources: WM task-positive coordinates were taken from the HCP-based WM-network activation analysis reported by Moser et al (Moser et al. 2018), who identified suprathreshold 2-back versus 0-back activations using random-effects one-sample t-tests with voxelwise family-wise error correction (FWE $p < 0.05$), and reported canonical bilateral dlPFC, inferior parietal, dACC and visual cortex peaks. DMN/task-negative coordinates were taken from Laird et al. (Laird et al. 2009) coordinate-based meta-analytic modeling of the default mode network (DMN).

# Pairwise maximum-entropy model (PMEM)

To conduct the SNM we transformed brain states into interpretable parameters. Firstly, we converted each regional signal to a binary state $s_i(t) \in -1, +1$ using a subject-independent threshold in units of the regional standard deviation: Schaefer-100: threshold (=0.8 s.d.); activation-derived 21-ROI system: threshold (=0.6 s.d.). At each time point ($t$), the brain state is $s(t) = [s_1(t), \ldots, s_N(t)]$. Based on a previous study, a broad range of thresholds can be selected for binarizing fMRI signals to obtain accurate fits of the pairwise maximum-entropy (Ising) model (PMEM) (Watanabe et al. 2013). Our recent analysis showed that a threshold near 1 s.d. can maintain the FC of the binarized time series as the original fMRI time series, achieve good model fitting, and can best reflect the biological regional heterogeneity of the brain (Craig et al. 2026). These threshold values were selected according to these criteria.

We modeled the empirical distribution of binary brain states with PMEM (Watanabe et al. 2013):

$$P(s;\theta) = \frac{1}{Z(\theta)} \exp\left(\beta\left(\sum_{i=1}^{N} h_i s_i + \sum_{i<j} J_{ij} s_i s_j\right)\right) \tag{1}$$

where $\theta = (h_i, J_{ij})$. contains regional effective excitability terms $h$ and interregional coupling terms $J$, and $Z(\theta)$ is the partition function. We parameterized $\theta$ as a vector of length $(M = N + N(N-1)/2)$ containing all $h_i$ and the upper-triangular $J_{ij}$ entries.

For each subject, we initialized the model at **(**$\beta = 1$**)** (Metropolis temperature parameter used in sampling during fitting). To accelerate convergence, we initialized parameters from the empirical moments of the binarized time series:

$$h_i^{(0)} \leftarrow \langle s_i \rangle_{data},$$

$$J_{ij}^{(0)} \leftarrow \mathrm{Cov}data(s_i, s_j) = \langle s_i\, s_j \rangle_{data} - \langle s_i \rangle_{data} \langle s_j \rangle_{data}. \tag{2}$$

We then updated parameters by gradient steps on the moment-matching objective:

$$h_i^{\mathrm{new}} = h_i^{old} - \alpha(\langle s_i \rangle_{model} - \langle s_i \rangle_{data}),$$
$$J_{ij}^{\mathrm{new}} = J_{ij}^{old} - \alpha\left(\langle s_i s_j \rangle_{model} - \langle s_i s_j \rangle_{data}\right), \tag{3}$$

with learning rate $\alpha = 10^{-3}$. Model expectations were estimated using Metropolis Monte Carlo 100,000 samplings. We quantified goodness-of-fit by the similarity between empirical and model-derived functional connectivity (FC), defined on binary states as

$$\mathrm{FC_{ij}} = \langle s_i s_j \rangle - \langle s_i \rangle \langle s_j \rangle \tag{4}$$

We assessed subject-level fit by comparing individual empirical binary-state FC with the FC simulated by the fitted PMEM, and we inspected fitting trajectories across participants (Supplementary Fig. S1A). For the cortex-only whole-brain system, subjects were retained if the fit reached a similarity criterion of 0.90, yielding n=1036 subjects. For the activation-derived 21-ROI system, subjects were retained based on an analogous convergence criterion of 0.95, a higher threshold used for smaller system as it is computationally feasible.

We then estimated a group PMEM for each system from concatenated binary states of the individual participants to obtain the group parameter $\overrightarrow{\theta_g}$. This group model served as the group reference where individual parameter deviations $\Delta\overrightarrow{\theta_q} = \overrightarrow{\theta_q} - \overrightarrow{\theta_g}$ is used to quantify individual difference in the brain networks.

# Fisher information matrix (FIM) and eigennetwork construction

After estimating the group PMEM, we computed the Fisher information matrix (FIM) to quantify how small, coordinated deviations in model parameters change the probability distribution of whole-brain states around the group reference model $\overrightarrow{\theta_g}$. Specifically, for a nearby parameter vector $\overrightarrow{\theta_g} + \overrightarrow{\delta\theta}$, the Kullback–Leibler (KL) divergence between the corresponding model distributions can be expanded locally around $\overrightarrow{\theta_g}$, such that the first-order term vanishes at the optimum and the second-order term is determined by the FIM. The FIM provides a local measure of parameter sensitivity, where directions with larger curvature indicate parameter combinations for which small perturbations lead to larger changes in the predicted distribution of brain states. In the context of PMEM, this framework has been used to distinguish "stiff" directions that strongly affect population activity from "sloppy" ones with minimal impact (Panas et al. 2015).

For the PMEM, the relevant sufficient statistics are the regional activity terms and pairwise co-activity terms, i.e. $s_i$ for the excitability parameters $h_i$ and $s_i s_j$ for the coupling parameters $J_{ij}$. Accordingly, the observed FIM is given by the covariance matrix of these sufficient statistics,

$$F_{ab} = \langle O_a O_b \rangle - \langle O_a \rangle \langle O_b \rangle, \tag{5}$$

where $O_a$ and $O_b$ denote the sufficient statistics associated with the *a-th* and *b-th* parameters in the vectorized PMEM parameterization. Thus, each entry of the FIM measures how strongly two parameter directions jointly affect the fitted distribution of brain states around the group model. This covariance-form expression follows directly from the exponential-family structure of the PMEM and is the form used in prior neural sloppiness analyses (Panas et al. 2015).

To improve numerical accuracy, we estimated these expectations directly from the empirical group-level binary time series, rather than via finite-step perturbation simulations around the fitted model. In practice, we concatenated the retained subjects' binarized time series to form the group dataset, computed the empirical sufficient statistics and their covariance, and used this covariance matrix as the group-level FIM. This plug-in estimator avoids the additional numerical noise introduced by repeated small-perturbation simulations while preserving the local sensitivity geometry around the empirically fitted group reference.

We then eigendecomposed the FIM,

$$F v_k = \lambda_k v_k, \tag{6}$$

obtaining eigenvalues $\lambda_k$ and eigenvectors $v_k$, ordered by decreasing eigenvalue (k=1, stiffest). Larger eigenvalues indicate parameter combinations along which small deviations from the group reference induce larger changes in the model's predicted state distribution. Each eigenvector therefore defines a sensitivity-ranked coordinated direction in PMEM parameter space, with its components specifying how strongly each $h_i$ and $J_{ij}$ contributes to that direction.

To visualize these directions as eigennetworks, each eigenvector $v_k$ was mapped back from the vectorized PMEM parameter space to an $N \times N$ symmetric matrix. The components corresponding to the regional excitability parameters $h_i$ were placed on the diagonal, whereas the components corresponding to the pairwise coupling parameters $J_{ij}$ were placed in the upper triangle and then mirrored to the lower triangle. In this matrix form, diagonal entries represent regional sensitivity weights and off-diagonal entries represent interregional sensitivity weights, allowing each FIM eigenvector to be expressed as an interpretable eigennetwork in matrix and graph form (Fig. 2E, F). This mapping preserves the one-to-one correspondence between the PMEM parameters and the entries of the eigennetwork representation.

Representability analyses across repeated data splits showed high overlap between fold-specific and full-sample leading FIM eigenspaces (Supplementary Fig. S1B), supporting the use of the full-sample group PMEM and its associated FIM as a stable reference for the main analyses.

## Optimal combination of eigennetworks to obtain behavior-aligned task-processing network and Behavior-AID

To link sensitivity geometry to behavior, we expressed each participant's deviation from the group PMEM in the eigennetwork basis and optimized a weighted linear combination of sensitivity-ranked eigennetworks to maximize out-of-sample association with a given task performance. This weighted pattern, the behavior-aligned direction $\overrightarrow{v_{\mathrm{op}}}$ in the parameter space (Fig. 2F), defines the task-processing network for a given task. The optimization was performed under repeated 3-fold cross-validation with progressive recruitment from stiff to sloppy eigennetworks. At each recruitment step, participant projections onto the currently recruited eigennetworks were entered into an Elastic Net model to evaluate whether the newly enlarged mode set improved held-out prediction of the target behavior. Predictability was quantified as the Pearson correlation between predicted and observed behavior in the held-out fold. Recruitment proceeded from stiff to sloppy eigennetworks, and the optimization was terminated according to the following stopping rule: if, after the addition of 50 consecutive eigennetwork modes, the improvement in test-set predictability was $< 0.01$, no further modes were recruited and the optimization was stopped.

To obtain a stable subject-level behavior-aligned axis for downstream prediction, the SNM training procedure was repeated 50 times, producing 50 training-derived $\overrightarrow{v_{\mathrm{op}}}$ estimates. These 50 vectors were then averaged to yield a single stabilized behavior-aligned direction, $\overline{\vec{v}_{\mathrm{op}}}$. For participant (q), the Behavior-AID was computed as the scalar projection of the individual parameter-deviation vector onto this averaged direction,

$$\mathrm{AID}^{(\mathrm{q})} = \Delta\overrightarrow{\theta_q} \cdot \overline{\vec{\mathrm{v}}_{\mathrm{op}}}. \quad (7)$$

After computing subject-level Behavior-AID values, behavioral prediction was performed using robust linear regression (MATLAB robustfit) on the training subjects and then applied to the held-out subjects. Predictability was quantified as the Pearson correlation between predicted and observed behavior in held-out folds.

Parameter sensitivity weights were defined as the absolute values of task-processing-network weights across both $h$ and $J$ components. We performed sequential removal analyses by progressively removing parameters from low-to-high sensitivity and from high-to-low sensitivity, recomputing Behavior-AID and its association with behavior after each removal level. To summarize degradation asymmetry, we computed the cumulative drop metric $D(n) = 1 - R(n)/R(0)$, where $R(n)$ is the correlation between Behavior-AID and behavior after removing $n\%$ of parameters. We further quantified coordinated-network integrity during sequential removal using giant component size in the binarized sensitivity-defined graph.

# Comparative models and stability analyses (Fig. 5-6)

For MVPA, we used a sklearn-based implementation. In this pipeline, each participant's regional BOLD time series ($X^{(q)} \in \mathbb{R}^{\mathbb{N}\times\mathbb{T}}$) was converted into a subject-level feature vector by averaging each ROI time series across time, yielding an (N)-dimensional feature vector. These features were z-scored across subjects using StandardScaler in sklearn, and prediction was performed with ridge regression. Model performance was evaluated in an outer 3-fold cross-validation. Thus, the MVPA baseline used distributed region-wise mean-activation features rather than connectivity or dynamical-parameter features.

For CPM, we used the YaleMRRC CPM implementation and followed the standard CPM workflow of feature selection, feature summarization, model fitting, and cross-validated prediction (Shen et al. 2017). Within each training fold, each FC edge was correlated with the target behavioral score across training subjects. Edges passing the predefined significance threshold were retained and separated into positively and negatively associated edge sets according to the sign of the correlation.

For each subject, the strengths of retained positive and negative networks were summarized by summing the selected edge weights, and these summary measures were then used to fit the predictive model and generate held-out predictions. Unless otherwise stated, the standard CPM analyses used the repository default threshold of ($p$<0.01); for Fig. 5B we additionally varied the CPM feature-selection threshold across a range of increasingly stringent (p)-values to examine the dependence of predictability on feature count.

To compare feature stability between CPM and SNM, we repeated the complete SNM workflow and the complete CPM workflow 10 times. For SNM, each run yielded one task-processing network $\overrightarrow{v_{op}}$; for CPM, each run yielded one selected-edge mask (or signed selected-edge set) in FC space. We then vectorized the resulting feature representations within each method and quantified within-method similarity across runs using pairwise cosine similarity. This produced 90 pairwise similarities per method ($10 \times 9/2$), which are shown in Fig. 6A.

For the sensitivity-aware CPM controls in Fig. 5, standard CPM was modified in three ways. In Enhanced CPM, the original CPM significance criterion was supplemented by an exemption rule: if an FC edge did not pass the CPM significance threshold but its SNM-derived sensitivity exceeded the preset sensitivity threshold (0.1 in Fig. 5), it was still retained as a predictive edge. In Sensitivity-only CPM, CPM was restricted to the subset of high-sensitivity edges, and statistical significance testing was performed only within this subset. In Sparse SNM, we used only the high-sensitivity edges, without CPM-style significance screening, and predicted behavior by 3-fold robust regression (robustfit) using the AID.

## Cross-modal alignment analyses (FC, SC and microstructure; Fig. 7)

To test whether the task-processing network provides a common behavior-aligned axis across modalities of functional connectivity, structural connectivity and microstructure, we constructed sensitivity-weighted alignment scores for participant-level FC, structural connectivity (SC) and mean diffusivity (MD).

Let $v_{op,ij}^{(J)}$ denote the coupling component of the task-processing network for edge i,j, and let $v_{op,i}^{(h)}$ denote its regional excitability component for node (i). For a subject ($q$), the FC alignment score was defined as

$$AID_{FC}^{(q)} = \sum_{i<j} v_{op,ij}^{(J)} FC_{ij}^{(q)} \tag{8}$$

that is, the upper-triangular sum of the element-wise product between the subject's FC matrix and the coupling-weight component of the task-processing network. $AID_{FC}^{(q)}$ thus measures the projection of individual FC along the task-processing network $\overrightarrow{v_{\mathrm{op}}}$.

Likewise, the MD alignment score was defined as

$$AID_{MD}^{(q)} = \sum_{i=1}^{N} v_{op,i}^{(h)} MD_i^{(q)} \tag{9}$$

where $MD_i^{(q)}$ is the regional mean diffusivity of subject ($q$). $AID_{MD}^{(q)}$ thus measures the projection of individual MD along the task-processing network $\overrightarrow{v_{\mathrm{op}}}$.

For SC, we used an analogous edge-wise weighted summary,

$$AID_{SC}^{(q)} = \sum_{i<j} v_{op,ij}^{(J)} SC_{ij}^{(q)} \quad (10)$$

Before computing this quantity, the diagonal of each SC matrix was removed (set to zero), because tractography-based self-connections correspond to streamlines beginning and ending within the same ROI are generally not treated as reliable inter-regional anatomical connections in structural connectome analyses; accordingly, self-connections are commonly removed and the connectome is analyzed with a zero diagonal. $AID_{SC}^{(q)}$ thus measures the projection of individual SC along the task-processing network $\overrightarrow{v_{\text{op}}}$.

For FC, the weighted summary was related to overall WM accuracy because FC was estimated from the full WM task time series (0-back + 2-back). For SC and MD, which were treated as more trait-like structural measures, the weighted summaries were computed using the task-processing network optimized for 2-back accuracy and were related to 2-back performance. Unweighted global FC and unweighted subnetwork-strength summaries were analyzed as controls.

# Transcriptomic association analysis and spin-test inference (Fig. 7 and Supplementary Fig. S8)

We related task-processing-network-derived regional maps to cortical gene-expression profiles from the Allen Human Brain Atlas processed with abagen (Markello et al. 2021) (probe selection via differential stability; donor-aggregated regional expression). We computed node-wise sensitivity strength maps separately for sensitivity positive and negative components of the task-processing network and tested gene-wise associations using Spearman correlation. To account for spatial autocorrelation, we assessed significance using an Alexander–Bloch spherical-rotation ("spin") null model (Alexander-Bloch et al. 2013), generating null maps by rotating cortical data on the sphere and recomputing correlations to form empirical spin-based p-values; we used 5,000 permutations where noted. Genes passing spin-test significance were ranked by effect size (absolute Spearman (ρ)); PANX1 was identified as the top-ranked gene under this procedure.

# Supplements:

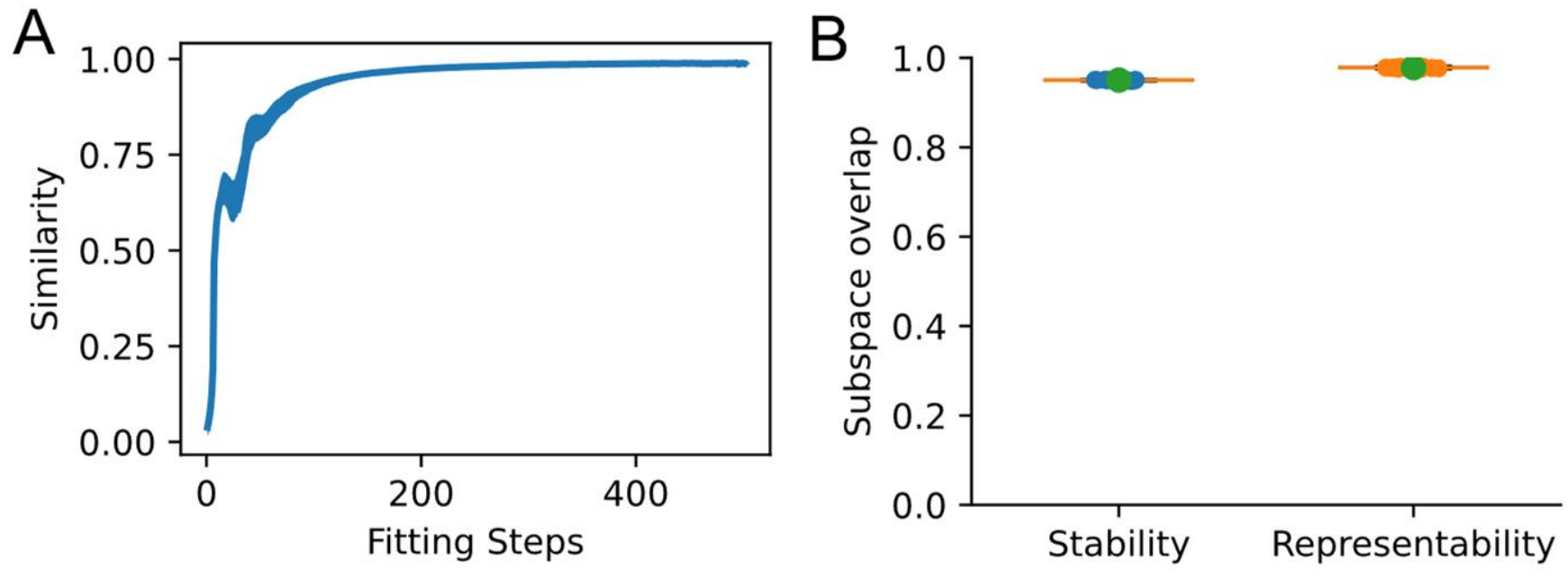

**Supplementary Fig. S1 |PMEM Fitting trajectories across subjects and stability and representability of the group-level FIM.**

**(A)** PMEM fitting trajectories for all retained subjects in the Schaefer-100 system. The y-axis shows the Pearson correlation between each subject's empirical binary-state functional connectivity (FC) and the FC simulated by the fitted PMEM; the x-axis shows fitting iterations. The shaded region indicates the interquartile range (25th–75th percentile) across subjects, illustrating stable convergence behavior and between-subject variability in fitting speed. **(B)** Stability and representability of Fisher-information eigenspaces under repeated 3-fold splits. For each split, we computed fold-specific group FIMs and the FIM from all concatenated subjects. We quantified overlap between the leading eigenspaces using the subspace-overlap metric $\text{Subspace Overlap}(U, V) = \frac{1}{r}|U\top V|_F 2$, where $U, V \in R^{M\times r}$ contain the top-(r) FIM eigenvectors (here $r$=300, matching the practical upper range of eigennetwork recruitment used in optimization in Supplementary Fig. S2). Repeating the 3-fold procedure 10 times, we report within-split fold-to-fold overlap (FIM stability) and overlap between the full-sample group FIM and each fold FIM (representability). High overlap indicates that the concatenated-data group FIM provides a robust and representative reference for the population-level sensitivity geometry. Each dot represents one pairwise similarity value from a specific split pair, and the solid line indicates the corresponding mean across all pairs.

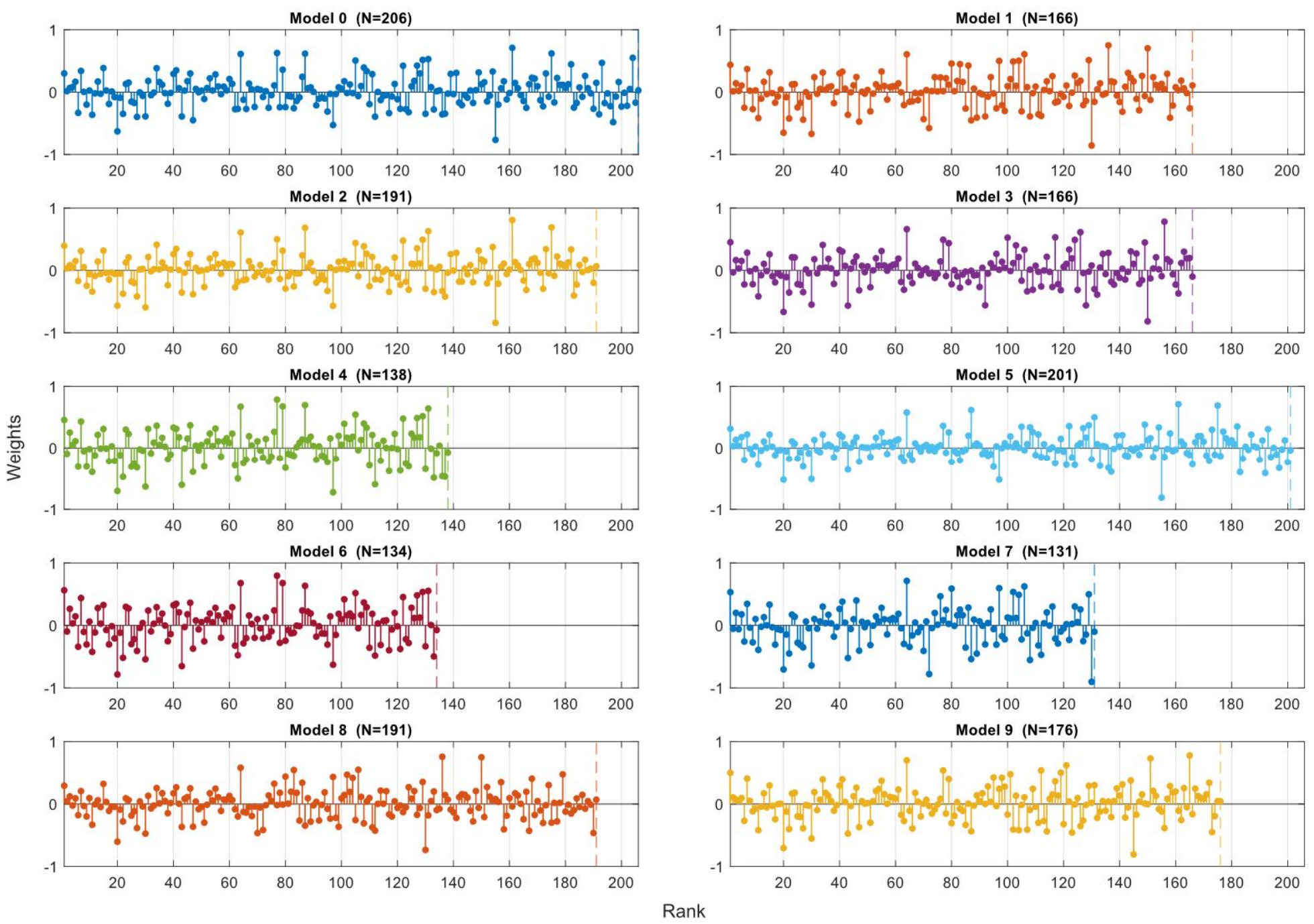

**Supplementary Fig. S2 | Mode recruitment weights show that SNM primarily uses stiff dimensions.**

To make explicit which eigennetwork dimensions are actually used by SNM after optimization, we visualized the recruitment weights assigned to eigennetwork modes. We repeated the full SNM workflow 10 times; within each run, the optimization itself was repeated 20 times and the resulting mode weights were averaged across the 20 repetitions to yield a single weight profile for that run. The plots therefore summarize, for each independent run, the mean recruitment weight of each mode. The sampling-averaged number of eigennetwork modes recruited (non-zero weight) is 165 ($s.d. \sim 32$), corresponding to only $\sim$ 3.2% of all 5050 eigennetworks.

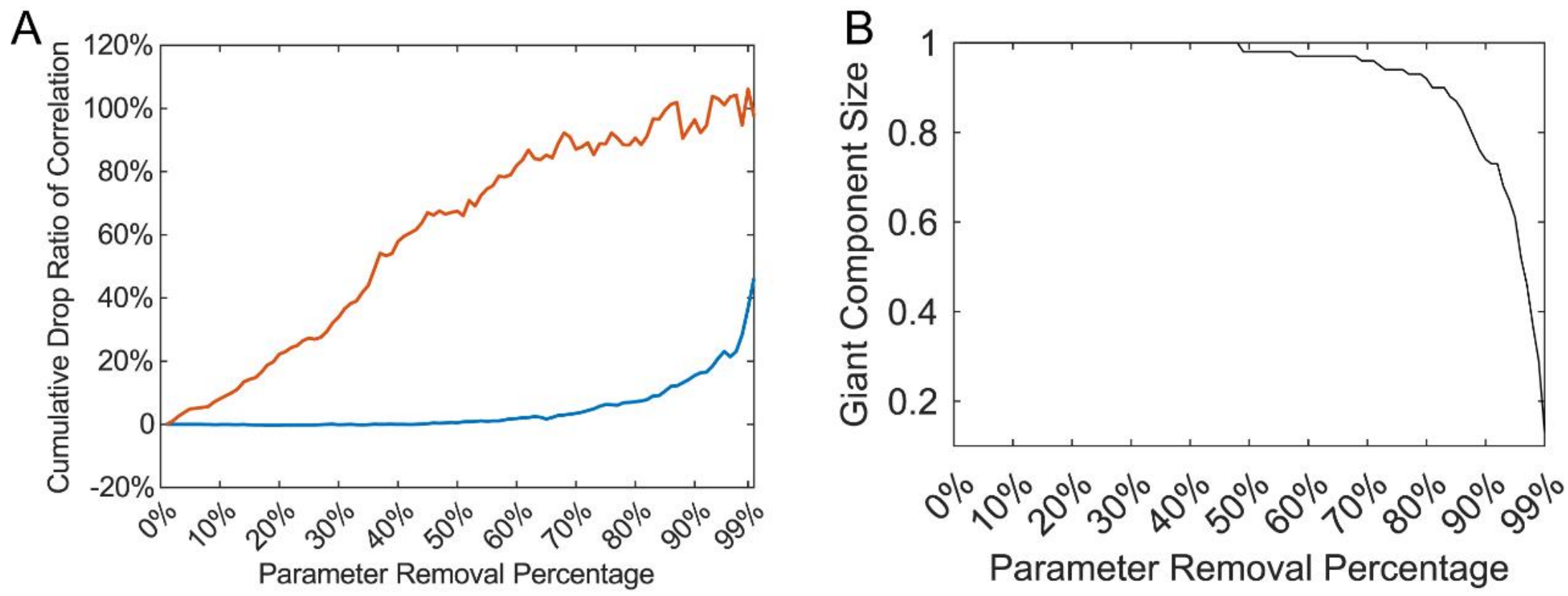

**Supplementary Fig. S3 | Quantifying sensitivity-dependent degradation by cumulative drop rate and coordinated-network integrity.**

To quantify the asymmetry in the loss of correlation between Behavior-AID and cognitive performance when removing high and low sensitivity parameters in the sparsefication of the SNM-derived cognitive processing network observed in Fig. 4A, we defined the **cumulative drop rate of correlation** as $D(n) = 1 - \frac{R(n)}{R(0)}$, where $(R(n))$ is the correlation between Behavior-AID and task performance after removing $(n\%)$ of parameters. **(A)** Cumulative drop-rate curves for removing **low-sensitivity (bleu)** versus **high-sensitivity (red)** parameters in the cortex-only whole-brain system. Removing insensitive parameters produces a slow increase in $D(n)$, with the cumulative drop rate reaching 20% only after approximately **95%** of parameters are removed. In contrast, removing sensitive parameters leads to a rapid increase, reaching 20% drop after removing approximately **20%** of parameters, quantitatively confirming the strong asymmetry implied by Fig. 4A. **(B)** Network-level structural change during sequential removal quantified by **giant component size**, GiantComponentSize $= \frac{|C_{max}|}{N}$, where $C_{max}$ is the largest connected component in the sparsified task-processing network and $|C_{max}|$ its number of nodes. The giant-component trajectory shows a trend closely matching the gradual-preservation regime of the low-sensitivity-removal correlation curve (Fig. 4A), indicating that Behavior AID–performance associations remain stable while the sensitivity-defined network retains a coordinated large-scale structure, and decline only after this structure is substantially disrupted.

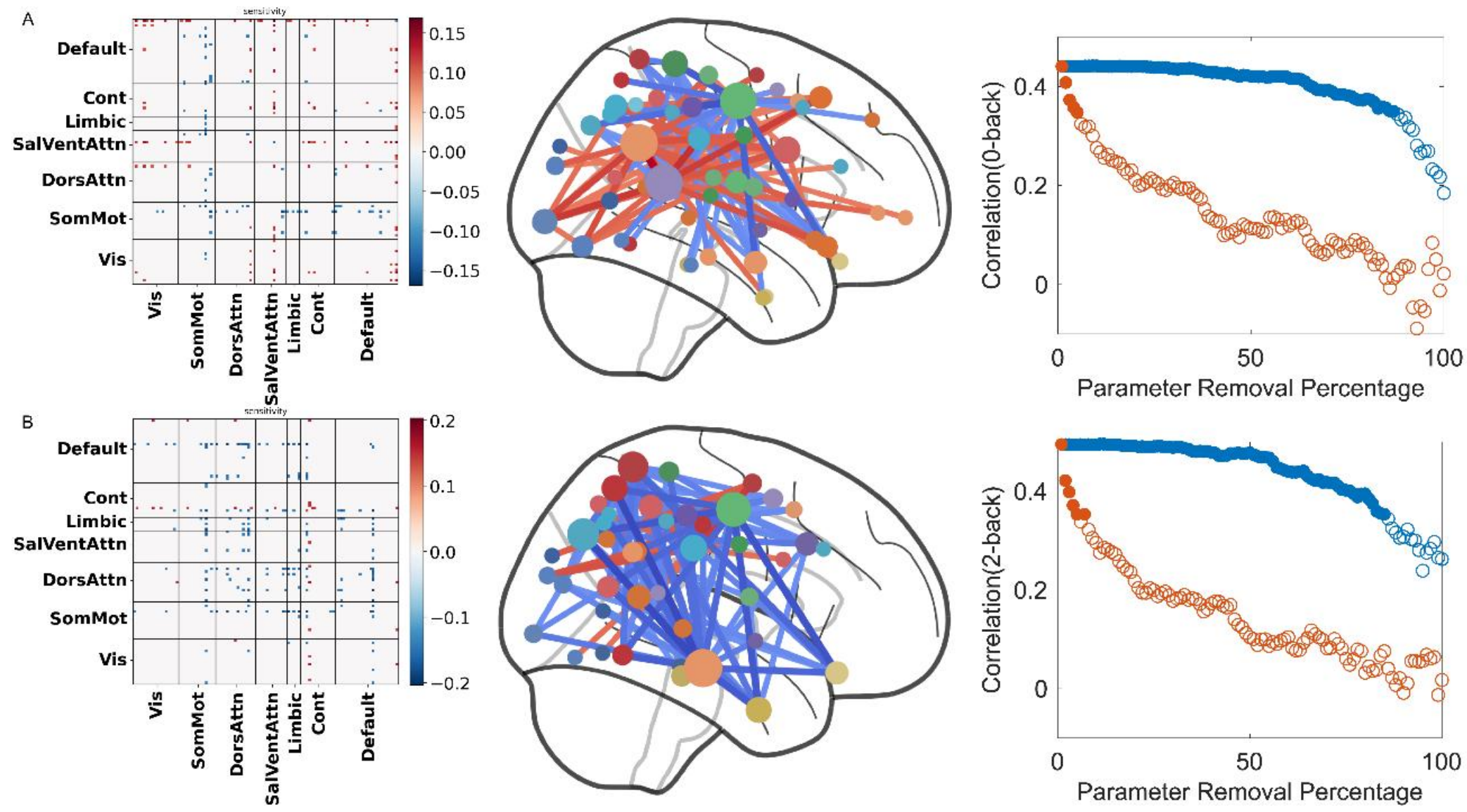

**Supplementary Fig. S4 | Condition-specific WM task-processing networks preserve sensitivity-dependent sparsification patterns.**

We repeated the sensitivity-removal analysis using task-processing networks (cortex-only whole-brain system)_optimized separately for 0-back and 2-back performance. For each target, we visualized the sparsified task-processing networks by retaining the most sensitive positive/negative edges (top 2%), and then quantified how the association between Behavior AID and performance score degrades as edges are removed either from high-to-low sensitivity or low-to-high sensitivity. In both conditions, removing high-sensitivity edges rapidly abolished the AID–performance association, whereas removing low-sensitivity edges preserved the association across a broad range of retained parameters, mirroring the sensitivity-dependent degradation pattern reported in the main analyses for the overall WM accuracy. The analysis also showed the cognitive processing network for 0-back and 2-back are distinct but have the common sparse and topologically coherent organization.

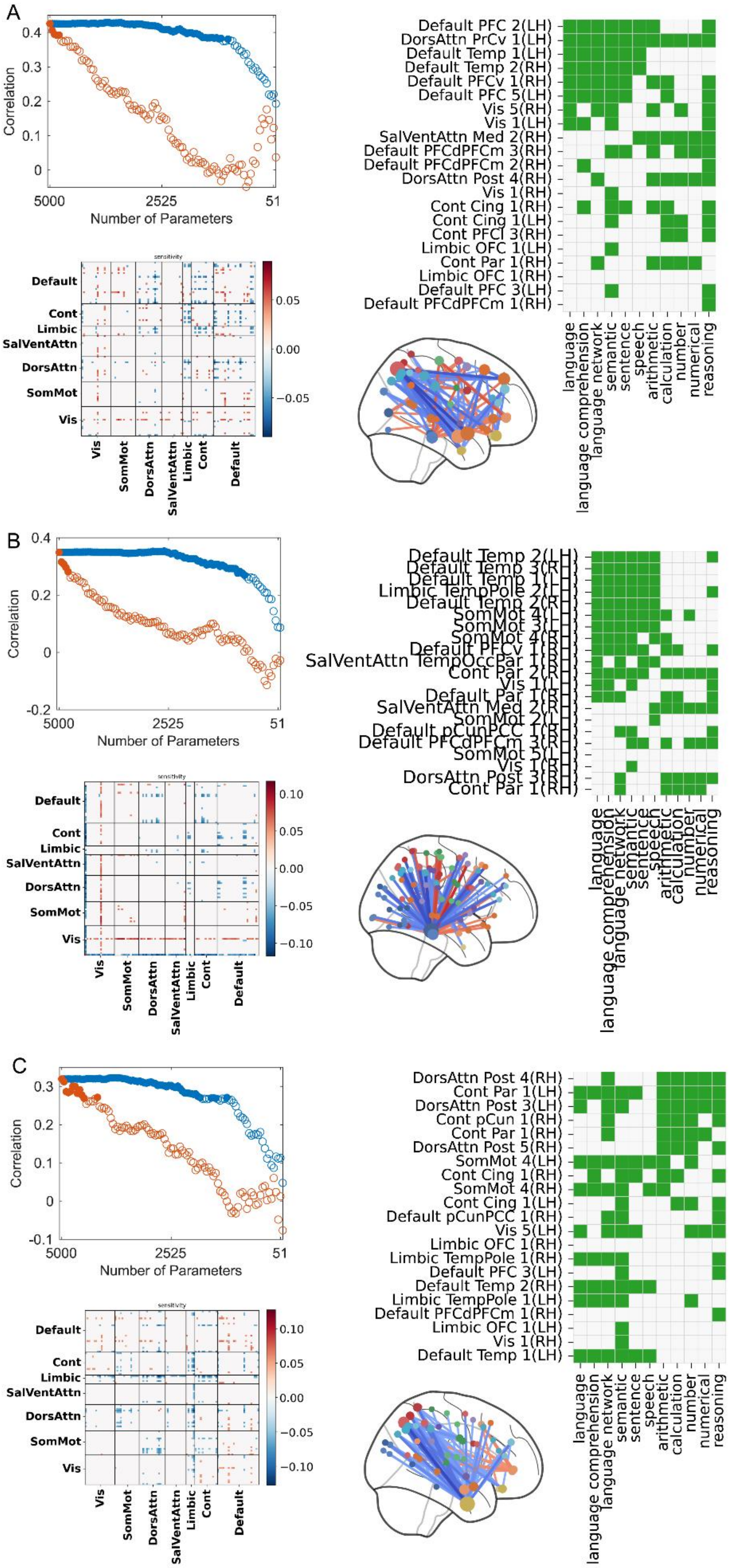

**Supplementary Fig. S5 | SNM generalizes to HCP language task and reveals target-specific sensitivity-defined subnetworks.**

We applied the same SNM pipeline used for working memory to the HCP language task, which interleaves auditory story-comprehension blocks and auditory math blocks across two runs (story and math conditions matched in block duration on average; see HCP task description). We optimized the task-processing network (cortex-only whole-brain system) using three language-performance targets: **(A)** overall language-task accuracy, **(B)** story-task accuracy, and **(C)** math-task accuracy. For each target, we performed sequential parameter-removal analyses of the SNM-identified task processing network analogous to Fig. 4 and Supplementary Fig. S4, removing parameters either from lowest-to-highest sensitivity or from highest-to-lowest sensitivity, and quantified the correlation between Behavior-AID and the corresponding accuracy score as a function of retained parameters. Across all three targets, removing low-sensitivity parameters preserved the AID–accuracy association across a broad range before gradual decline, whereas removing high-sensitivity parameters caused a rapid loss of association, consistent with the sensitivity-dependent degradation pattern observed in working memory. Each panel shows the top-21 sensitivity-defined regions (ranked by absolute regional sensitivity) and a term-by-region Neurosynth meta-analytic association matrix using language- and numerosity-related terms (language, language comprehension, language network, semantic, sentence, speech, arithmetic, calculation, number, numerical, reasoning); green cells indicate positive associations. The corresponding sparsified sensitivity matrix and whole-brain network rendering (after removing **95%** low-sensitivity parameters) visualize the sparse and topologically coherent language-task-processing subnetwork. Although each analysis selects the top-21 most sensitive regions, the resulting subnetworks and meta-analytic profiles differ across overall/story/math targets, indicating that SNM resolves target-specific functional specialization within the language-task paradigm.

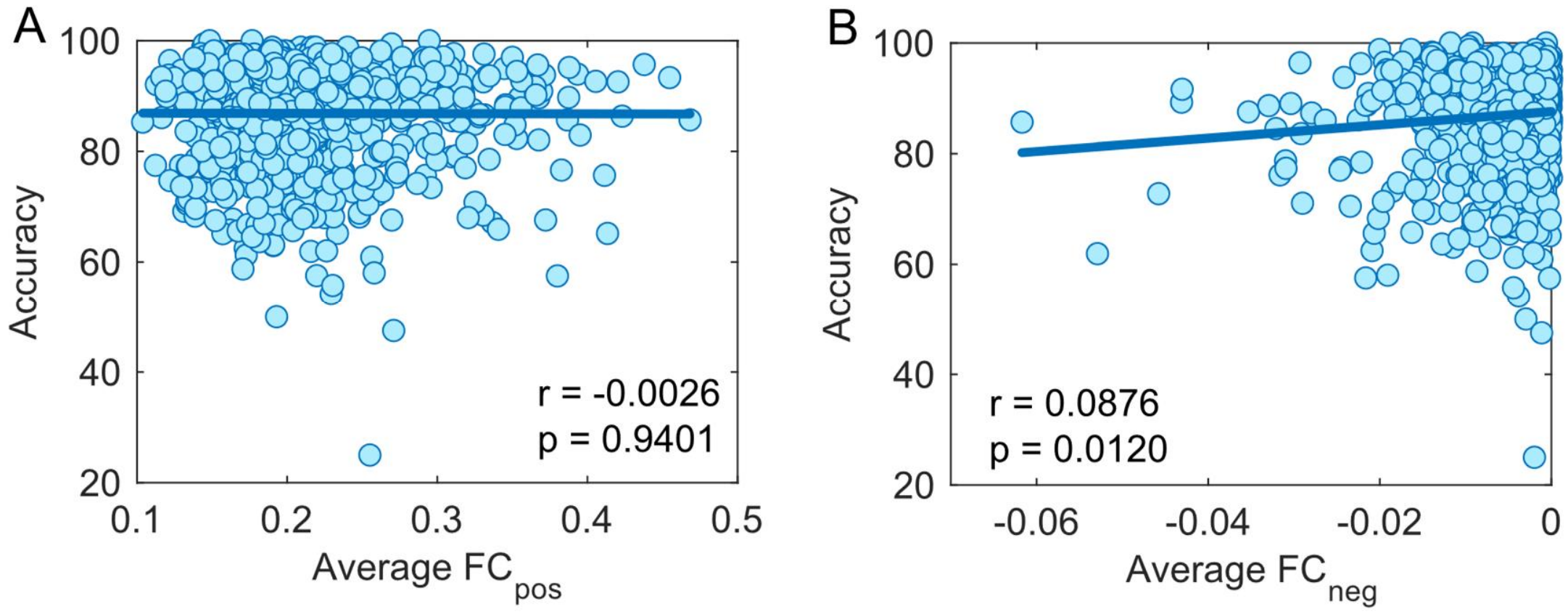

**Supplementary Fig. S6 | Mean strength of sensitivity positive/negative subnetworks is not sufficient to explain behavior.**

We tested whether simple strength summaries of the sensitivity positive and negative FC subnetworks capture individual differences in overall WM accuracy. After splitting FC into positive and negative components, we computed each participant's mean FC in positive subnetwork and and negative subnetwork and correlated them with overall WM accuracy. Neither measure showed a significant association, indicating that global magnitude summaries within sensitivity sign-defined subnetworks are not, by themselves, behavior-informative**.**

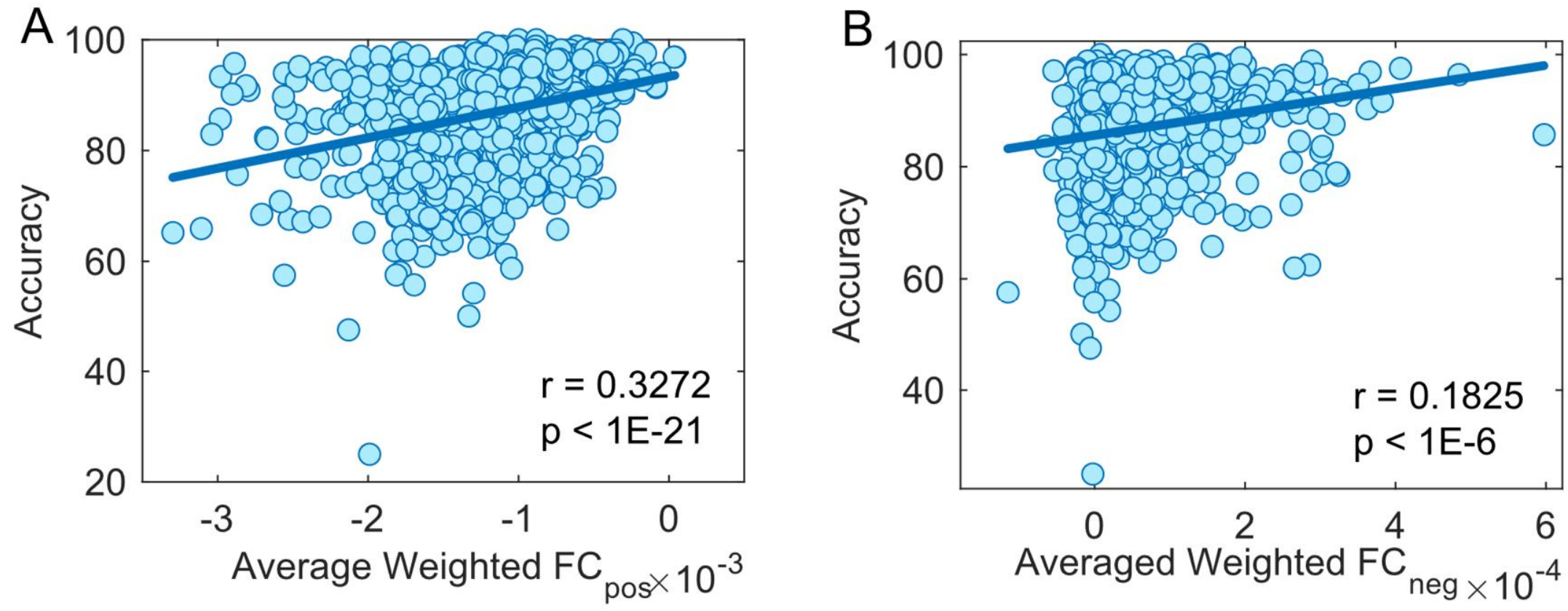

**Supplementary Fig. S7 | Sensitivity-weighted FC strength reveals behavior-relevant signal in sensitivity positive and negative subnetworks.**

We next asked whether the same sensitivity positive/negative subnetworks become behavior-relevant once they are evaluated along the task-processing network, i.e., with weights interpreted as parameter sensitivity. For each participant, we computed sensitivity-weighted FC strength by weighting FC edges using the task-processing network and summing within sensitivity positive and negative subnetworks, respectively. In contrast to unweighted strength, sensitivity-weighted strength showed robust associations with overall WM accuracy. These results reinforce the main conclusion that behavioral signal is expressed through sensitivity-aligned coordination, rather than through undirected changes in overall FC magnitude.

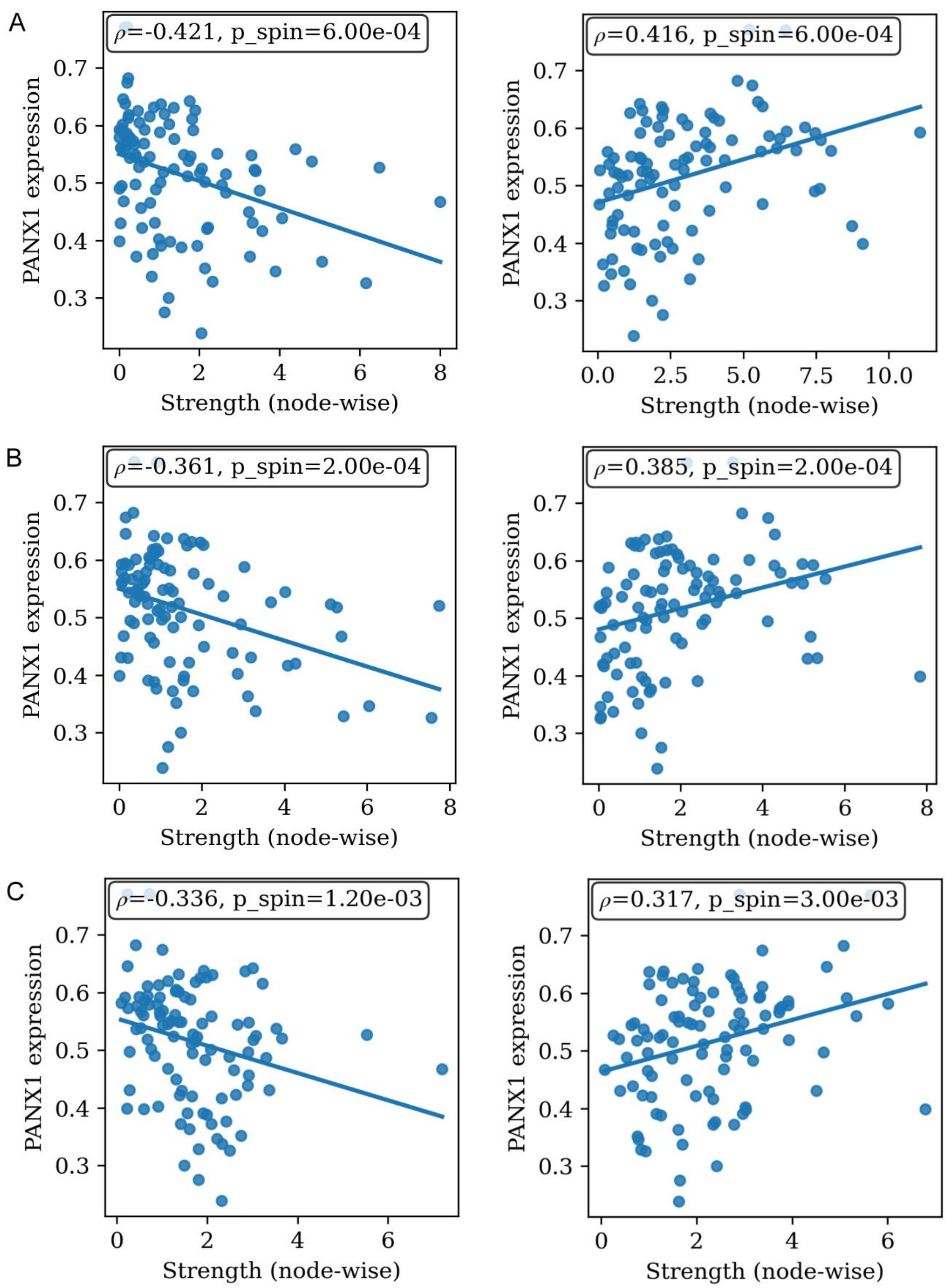

**Supplementary Fig. S8 | PANX1 associations generalize across task targets and sign-defined subnetworks.** To evaluate whether the transcriptomic association with working memory observed in the main text depends on the specific behavioral target used for optimization, we derived task-processing networks optimized separately for 2-back accuracy, 0-back accuracy, and overall accuracy, then split each task-processing network into sensitivity positive and negative parts

and computed region-wise strength maps. Across optimization targets, regional strength maps for both subnetworks showed significant associations with PANX1 expression, with significance assessed using spatially constrained spin tests (spin-test threshold ($p<0.01$)). Together, these analyses indicate that association between PANX1 and working memory is not an idiosyncratic result of a single optimization target, but is consistently linked to the sensitivity-defined architecture captured by the task-processing network.

Supplementarty Table. 1| Coordinates of sphere ROI definition of the activation-derived system. All coordinates are unified to MNI space using tal2mni of nimare.

| -2.60318390264692 | -56.4178069850627 | 49.7623863939712 |
|---|---|---|
| -2.80865278686680 | -51.9579470401397 | 24.8563251267127 |
| 3.02258359382131 | 34.3327690778987 | -15.9032640188183 |
| 56.9544558518304 | -26.0503141836157 | 24.2275596423933 |
| -1.10833037054496 | 55.5677176731790 | 11.3585629907367 |
| 50.6272269345471 | -66.9554965478582 | 18.8090698148468 |
| -26.4053692128873 | 21.6894538349418 | 43.4854782809246 |
| -58.3896512857136 | -34.8583885899690 | 30.7519395898389 |
| -43.3930696437716 | -67.3475908368624 | 22.1523223466744 |
| 44 | -38 | 42 |
| -42 | -48 | 44 |
| 28 | 8 | 58 |
| -28 | 4 | 60 |
| 2 | 22 | 44 |
| -2 | 24 | 44 |
| 4 | 12 | 26 |
| -12 | -94 | 0 |
| 16 | -90 | -8 |
| -28 | -86 | -16 |
| -50 | -38 | -6 |
| 50 | -42 | -2 |

## Acknowledgement

This work was partially supported by STI 2030-Major Projects (No. 2022ZD0208500), the Hong Kong Research Grant Council Senior Research Fellow Scheme (SRFS2324-2S05) and General Competitive Fund (GRF12202124), Guangdong and Hong Kong Universities "1+1+1" Joint Research Collaboration Scheme (2025A0505000011) and Hong Kong Baptist University (HKBU) Seed Funding for Collaborative Research Grants (RC-SFCRG/23-24/SCI/06). Data were provided by the Human Connectome Project, WU-Minn Consortium (Principal Investigators: David Van Essen and Kamil Ugurbil; 1U54MH091657) funded by the 16 NIH Institutes and Centers that support the NIH Blueprint for Neuroscience Research; and by the McDonnell Center for Systems Neuroscience at Washington University.